\documentclass[showpacs,aps,prstab,twocolumn,amsmath,amssymb]{revtex4-1}
\usepackage{amsmath}
\usepackage{graphicx}
\usepackage{subfig}
\usepackage{bm}
\usepackage{booktabs}
\usepackage[usenames]{color}
\usepackage{hyphenat}
\usepackage[titletoc,title]{appendix}
\usepackage{titlesec}
\usepackage{enumitem}

\titlespacing\section{0pt}{12pt plus 0pt minus 2pt}{0pt plus 0pt minus 0pt}
\titlespacing\subsection{0pt}{12pt plus 4pt minus 2pt}{0pt plus 2pt minus 2pt}
\titlespacing\subsubsection{0pt}{12pt plus 4pt minus 2pt}{0pt plus 2pt minus 2pt}

\renewcommand{\thesubsection}{\thesection\Alph{subsection}}

\begin{document}
\def\thesection{\arabic{section}}
\numberwithin{equation}{section}


\title{Analysis of Nonlinear Dynamics by Square Matrix Method }
\author{Li Hua Yu}

\affiliation{%
Brookhaven National Laboratory, Upton, NY 11973
}%

\begin{abstract}
 The nonlinear dynamics of a system with periodic structure can be analyzed using a square matrix. We show that because the special property of the square matrix constructed for nonlinear dynamics, we can reduce the dimension of the matrix from the original large number for high order calculation to low dimension in the first step of the analysis. Then a stable Jordan decomposition is obtained with much lower dimension. The Jordan decomposition leads to a transformation to a new variable, which is an accurate action-angle variable, in good agreement with trajectories and tune obtained from tracking. And more importantly, the deviation from constancy of the new action-angle variable provides a measure of the stability of the phase space trajectories and tune fluctuation.  Thus the square matrix theory shows a good potential in theoretical understanding of a complicated dynamical system to guide the optimization of dynamical apertures. The method is illustrated by many examples of comparison between theory and numerical simulation. In particular, we show that the square matrix method can be used for fast optimization to reduce the nonlinearity of a system.
\end{abstract}


\maketitle

\section{Introduction}
\label{sec:Introduction}
The question of the long term behavior of charged particles in storage rings has a long history. To gain understanding, one would like to analyze particle behavior under many iterations of the one turn map. The most reliable numerical approach is the use of a tracking code with appropriate integration methods. For analysis, however, one would like a more compact representation of the one turn map out of which to extract relevant information. Among the many approaches to this issue we may mention canonical perturbation theory, Lie operators, power series, and normal form \cite{lieb_1983,ruth_1986,guinard_1978,shoch_1957,Dragt:1988wto,berz_1989,
chao_95,Turchetti_1994,forest_1989,miche_1995,forest_1998}, etc.

Here, we would like to look at this problem from a somewhat different perspective: we shall analyze the problem by the method of square matrix \cite{yu_2009,yu_2016}, constructed out of the power series map \cite{berz_1989}. In this paper, we first outline the concept of the square matrix method, then we discuss its application, and then explain the mathematical details of the method, about how and why it is suitable for high order calculation. In the following introduction, we first outline the basic concept of the square matrix analysis in subsection 1A-1D, then in subsection 1E we present the outline of the paper.

\subsection{Representation of nonlinear maps using square matrices}

We consider the equations of motion of a nonlinear dynamic system with periodic structure such as Hill’s equation, it can be expressed by a square matrix.

If we use the complex Courant-Snyder variable $z=x-i p$, its conjugate and powers $z,z^\ast,z^2,...$  as a column $Z$, the one turn map can be represented by a large square matrix $M$ using $Z=MZ_0$.

For example, for a simple Henon map \cite{Turchetti_1994},
\begin{equation}
   \label{eq:Eq.1.1}
   \begin{split}
  x=x_0 \cos \mu + p_0 \sin \mu + \epsilon x_0^2 \sin \mu
  \\
    p=-x_0\sin \mu + p_0 \cos \mu + \epsilon x_0^2 \cos \mu
  \end{split},
\end{equation}
we use the variables $z=x-i p$ and $z^ \ast =x+ip$ to write $z,z^\ast$ and their higher power monomials after one turn of rotation, or after one element in an accelerator lattice, as a truncated power series expansion of the initial $z_0=x_0 -i p_0$ and $z_0^ \ast =x_0+ip_0$. For example, up to 3rd order, we have:
{\small
\begin{equation}
   \label{eq:Eq.1.2}
\begin{split}
&z=e^{i \mu}z_0 - \frac{i}{4}\epsilon e^{i\mu} z_0^2 - \frac{i}{2}\epsilon e^{i\mu} z_0z_0^\ast - \frac{i}{4}\epsilon e^{i\mu} z_0^{\ast2}
  \\
&z^\ast=e^{-i \mu}z_0^\ast + \frac{i}{4}\epsilon e^{-i\mu} z_0^2 + \frac{i}{2}\epsilon e^{-i\mu} z_0z_0^\ast + \frac{i}{4}\epsilon e^{-i\mu} z_0^{\ast2}
  \\
&z^2=e^{2i \mu}z_0^2 - \frac{i}{2}\epsilon e^{2i\mu} z_0^3 - i\epsilon e^{2i\mu} z_0^2z_0^\ast - \frac{i}{2}\epsilon e^{2i\mu} z_0z_0^{\ast2}
  \\
&zz^\ast=z_0z_0^\ast + \frac{i}{4}\epsilon  z_0^3 + \frac{i}{4}\epsilon z_0^2z_0^\ast - \frac{i}{4}\epsilon z_0z_0^{\ast2}- \frac{i}{4}\epsilon z_0^{\ast3}
  \\
&z^{\ast2}=e^{-2i \mu}z_0^{\ast2} + \frac{i\epsilon  }{2}e^{-2i\mu} z_0^2z_0^\ast  + i\epsilon e^{-2i\mu} z_0z_0^{\ast2} + \frac{i\epsilon}{2} e^{-2i\mu} z_0^{\ast3}
  \\
&z^3=e^{3i \mu}z_0^3
    \\
&...
    \\
&z^{\ast3}=e^{-3i \mu}z_0^{\ast3}
  \end{split},
\end{equation}
}
In general, there are constant terms in the expansion, even though in this example, the offset of x is zero, so the constant terms are also zeros. We may now write this in the matrix form:
\begin{equation}
   \label{eq:Eq.1.3}
Z=MZ_0,
\end{equation}
where to 3rd order, we define the 10$\times$1 monomial array
\begin{equation}
   \label{eq:Eq.1.4}
\tilde{Z}=(1,z,z^\ast,z^2,zz^\ast,z^{\ast2},z^3,z^2z^\ast,zz^{\ast2},z^{\ast3}).
\end{equation}
The row $\tilde{Z}$ represents the matrix transposition of the column $Z$. The vector $Z$ spans a 10 dimensional linear space. The matrix $M$, when operated on $Z_0$ , represents a rotation to $Z$ in this space. We remark here that even though we mostly use $M$ to represent one turn map for a storage ring, it can be used to represent one element in the storage ring dynamics or other nonlinear dynamics problem.

The square matrix $M$ is upper-triangular, and has the form:
\begin{equation}
   \label{eq:Eq.1.5}
   M=
\begin{bmatrix}
  1 & 0 & 0 & 0 \\
  0 & M_{11} & M_{12} & M_{13} \\
  0 & 0 & M_{22} & M_{23} \\
  0 & 0 & 0 & M_{33}
 \end{bmatrix}
\end{equation}

The zeros here represent sub-matrixes with all zero elements. $M_{11}$, $M_{22}$ and $M_{33}$ are diagonal matrixes with dimension 2$\times$2, 3$\times$3 and 4$\times$4 respectively, their diagonal elements are $\{ e^{i\mu},e^{-i\mu}\}$,\{$e^{2i\mu},1 ,e^{-2i\mu}$\},\{$e^{3i\mu},e^{i\mu},e^{-i\mu},e^{-3i\mu}$\}. They correspond to 1st, 2nd, and 3rd order terms in the series expansion respectively, i.e., length 2, 3 and 4. The eigenvalues of a triangular matrix are its diagonal elements, hence, these 9 numbers and the first diagonal element 1, are the 10 eigenvalues of $M$. The sub-matrixes $M_{12},M_{13}$,and $M_{23}$ have dimension $2\times3$, $2\times4$, and $3\times4$ respectively, represent the cross terms between 1st order and 2nd order, 1st and 3rd order, and between 2nd order and 3rd order respectively. As an example, from Eq.(1.2), we find
\begin{equation}
   \label{eq:Eq.1.6}
   \begin{split}
      M_{11}&=
   \begin{bmatrix}
 e^{i\mu} &  0 \\
 0& e^{-i\mu}
 \end{bmatrix}, M_{22}=
 \begin{bmatrix}
 e^{2i\mu} &  0 & 0\\
 0 &  1 & 0\\
 0& 0& e^{-2i\mu}
 \end{bmatrix},...\\
   M_{23}&=
\begin{bmatrix}
- \frac{i}{2}\epsilon e^{2i\mu} & - i\epsilon e^{2i\mu} & - \frac{i}{2}\epsilon e^{2i\mu} &  0 \\
 \frac{i}{4}\epsilon  & \frac{i}{4}\epsilon & - \frac{i}{4}\epsilon &- \frac{i}{4}\epsilon \\
  0 &\frac{i}{2}\epsilon e^{-2i\mu} & i\epsilon e^{-2i\mu} &  \frac{i}{2}\epsilon e^{-2i\mu}
 \end{bmatrix}
\end{split}
\end{equation}

\subsection{Eigenvectors of Jordan blocks}
We would like to extract the spectral structure from the
matrix M. The first inclination may be to try to diagonalize
and find the eigenvalues. It turns out, however, that
for non-linear maps M is not diagonalizable.
All square matrices may be transformed into Jordan form,
however. Doing so, we find a transformation matrix $U$ and a Jordan matrix $\tau$ so that
every row of the matrix $U$ is a (generalized) left eigenvector of $M$ satisfying
\begin{equation}
   \label{eq:Eq.1.7}
   \begin{split}
UM =e^{i \mu I +{\tau}}U \\
   \end{split}
\end{equation}

In the above example, for the case of two variables $z,z^\ast$  at 3rd roder, the matrix $M$ is a $10 \times 10$ matrix, $I$ is a $2 \times 2$ identity matrix, the matrix $U$ is a $2 \times 10$ transformtion matrix, while the Jordan matrix
\begin{equation}
   \label{eq:Eq.1.8}
   \tau=
\begin{bmatrix}
 0    & 1   \\
 0    & 0
 \end{bmatrix}.
\end{equation}
In the general case, the Jordan matrix $\tau$ always has much lower dimension than the mapping matrix $M$, and has the form
\begin{equation}
   \label{eq:Eq.1.9}
   \tau=
\begin{bmatrix}
 0    & 1  & 0 &...& 0 \\
 0    & 0  & 1 &...& 0 \\
0    & 0  & ...&...& 0 \\
0    & 0  & 0&...& 1 \\
0    & 0  & 0 &...& 0
 \end{bmatrix}.
\end{equation}

In the  example for the case of 4 variables $x,p_x,y,p_y$  at 7'th order, as we shall explain later, the matrix $M$ is a $330 \times 330$ matrix, $I$ is a $4 \times 4$ identity matrix, the matrix $U$ is a $4 \times 330$ transformation matrix, while the Jordan matrix $\tau$ also has dimension 4.

As $Z=MZ_0$  (see Eq.\,\eqref{eq:Eq.1.3}), Eq.\,\eqref{eq:Eq.1.7} gives
\begin{equation}
   \label{eq:Eq.1.10}
UZ=U M Z_0= e^{i\mu I +{\tau}}UZ_0.
  \end{equation}

Now we define a transformation
\begin{equation}
   \label{eq:Eq.1.11}
   \begin{split}
W &\equiv UZ \\
W_0 &\equiv UZ_0 \\
   \end{split}
\end{equation}

$W$ represents the projection of the vector $Z$ onto the invariant subspace spanned by the left eigenvectors $u_j$ given by the rows of the matrix $U$, such that each row of $W$ is $w_j=u_j Z$, a polynomial of $z,z^\ast$. Then Eq.\,\eqref{eq:Eq.1.10} implies the operation of one turn map $Z= M Z_0$, corresponds to a rotation in the invariant subspace represented by
\begin{equation}
   \label{eq:Eq.1.12}
W= e^{i\mu I +{\tau}} W_0.
  \end{equation}

\subsection{Multi-turns behavior, coherent state, and frequency fluctuation}
The new vector after n turns becomes

\begin{equation}
   \label{eq:Eq.1.13}
W= e^{n(i\mu I +{\tau})} W_0=e^{i n \mu} e^{n \tau} W_0.
  \end{equation}

KAM theory states that the invariant tori are stable under small perturbation (See, for example, \cite{lieb_1983, Turchetti_1994, poschel_2001}). In our examples, for sufficiently small amplitude of oscillation in z, the invariant tori are deformed and survive, i.e., the motion is quasiperiodic. So the system has a nearly stable frequency, and when the amplitude is small, the fluctuation of the frequency is also small. Thus for a specific initial condition described by $Z_0$, the rotation in the eigenspace should be represented by a phase factor $e^{i(\mu+\phi)}$ so that after n turns
\begin{equation}
   \label{eq:Eq.1.14}
W= e^{n(i\mu I +{\tau})} W_0 \cong e^{in (\mu+\phi)} W_0.
  \end{equation}

We remark that if $W_0$ is an eigenvector of $\tau$ with eigenvalue of $i \phi$, i.e.,
\begin{equation}
\label{eq:Eq.1.15}
\tau W_0 \cong i \phi W_0.
\end{equation}
then Eq.(1.14) is satisfied.

We may rephrase this as: if $W_0$ is  a coherent state \cite{glauber_1963,sudarshan_1963} of $\tau$ with eigenvalue $i \phi$, then $\phi$ is the amplitude dependent tune shift.

In Eq.\,\eqref{eq:Eq.1.14} and Eq.\,\eqref{eq:Eq.1.15} we use the approximate equal sign because for a matrix $\tau $ of finite dimension m, there is only approximate coherent state, the coherent state exists only when the dimension m approaches infinity. To see this we write the Eq.\,\eqref{eq:Eq.1.14} explicitly using the property of the Jordan matrix $\tau$ given by Eq.\,\eqref{eq:Eq.1.9} as a lowering operator:
\begin{equation}
   \label{eq:Eq.1.16}
  \tau
  \begin{bmatrix}
  w_0\\w_1\\...\\w_{m-1}
   \end{bmatrix}=
  \begin{bmatrix}
  w_1\\w_2\\...\\0
   \end{bmatrix} \cong
\begin{bmatrix}
i\phi w_0\\i\phi w_1\\...\\i\phi w_{m-1}
 \end{bmatrix}
\end{equation}
where $w_j$'s are the rows of $W_0$. Compare the two sides we find
\begin{equation}
\label{eq:Eq.1.17}
i\phi  =  \frac{w_1}{w_0}\cong  \frac{w_2}{w_1}\cong \frac{w_3}{w_2} \cong \dots \cong \frac{w_{m-1}}{w_{m-2}}
  \end{equation}
It is obvious that the equal sign holds only as the dimension m approaches infinity.
In the study of truncated power series, m is finite, hence Eq.\,\eqref{eq:Eq.1.17} is an approximation. As we shall show later, the polynomials $w_{m-2},w_{m-1}$ have only high order terms, and as m increases, when the amplitude of $z$ is sufficiently small, the last term in Eq.\,\eqref{eq:Eq.1.17} becomes the ratio of two negligibly small numbers, and is less accurate.

In addition to the condition Eq.1.17 for a stable motion, obviously, another condition is
\begin{equation}
\label{eq:Eq.1.18}
\text{Im}(\phi) \cong 0.
  \end{equation}
The Eq.\,\eqref{eq:Eq.1.17} also shows that $w_1^2 \cong w_0 w_2$, and hence we have
\begin{equation}
\label{eq:Eq.1.19}
 \Delta \equiv \frac{w_2}{w_0}-(\frac{w_1}{w_0})^2 \cong 0.
  \end{equation}
Using Eq.(1.13), we can derive the multi-turn behavior of $w_0$ as an expansion of number of turns (See Section 7)
\begin{equation}
   \label{eq:Eq.1.20}
   \begin{split}
w_0(n)&= e^{ in\mu+ i n \phi+\frac{n^2}{2} \Delta +\dots}w_0(n=0).
  \end{split}
\end{equation}
The contribution to phase advance from $\Delta$ is proportional to $n^2$ instead of $n$. Hence the deviation of $\Delta$ from zero gives the information about the fluctuation of the frequency (tune variation) during the motion and the loss of stability of the trajectory. This seems to be related to the Liapunov exponent\cite{lieb_1983}( p.298), and hence we use it to find the dynamic aperture in the examples in the later sections.

The Eq.\,\eqref{eq:Eq.1.14}, derived for $x,y$ planes separately, leads to a set of action-angle variables $w_{0x};w_{0y}$, with its action
amplitude and phase advance angle nearly constant of
motion up to near the border of the dynamic aperture or
resonance lines. In addition, we find that near this border
the deviation of these actions from constancy provides a
measure of the destruction of invariant tori, or a measure
of the stability of trajectories and tunes.

We consider the functions $w_{0x}(z_x,z_y),w_{0y}(z_x,z_y)$ as a definition for a variable transformation. These functions and their inverse functions $z_x(w_{0x},w_{0y}),z_y(w_{0x},w_{0y})$ provide a one to one correspondence between the $z_x,z_y$ planes and the $w_{0x},w_{0y}$ planes. During the motion $|w_{0x}|,|w_{0y}|$ are only approximately constant, we do not use $w_{0x},w_{0y}$ for long term tracking of particles. However,the deviation of their amplitude from constant, described by the function $\Delta$ of Eq.1.19 in the $w_{0x},w_{0y}$ planes, provides a provides a measure of nonlinearity.

To clarify the relation between this variable transformation and the one turn map, we discuss the relation between four maps. The one turn map in $z_x,z_y$ (the first map) is given by the first two rows of Eq.(1.2). The one turn map of $w_{0x},w_{0y}$ (the second map) is exactly equivalent to the one turn map in $z_x,z_y$ planes, as long as the inverse function $z_x(w_{0x},w_{0y}),z_y(w_{0x},w_{0y})$ exists, because it is the same map expressed in terms of another set of variables. For sufficient high order, the non-existence of the inverse function in a region indicates the motion is unstable in that region, i.e, it is at the border of dynamic aperture. We do not have explicit expression for this second map, we only know it as an implicit function through the first map. The third map Eq.(1.12) is equivalent to $Z \cong MZ_0$, only provides an approximation to the first two exact maps. It is a truncated map, not symplectic. The fourth map is given by $W \cong e^{in (\mu+\phi)} W_0$. When we take $\phi$ to be a real constant determined from Eq.(1.17) by $W_0$, this map is a twist map\cite{lieb_1983,Turchetti_1994}.

The advantage of using the variables $w_{0x},w_{0y}$ is that the exact map in $w_{0x},w_{0y}$ only has a small deviation from the twist map (the four'th map). Hence the deviation is considered as a perturbation to the twist map. The deviation is embodied by the fact that, for the second map, $|w_{0}|$ and $\phi$ are not constant after one turn, and $\Delta \neq 0, \text{Im}(\phi) \neq 0$. Because the deviation is small even close to the dynamic aperture, the perturbed twist map can be used for analysis of long term behaviour, for example, as used in Poincare-Birkhoff theorem\cite{Turchetti_1994}. Similarly, the third map is also a perturbed twist map when we use $w_{0x},w_{0y}$ variables. The difference from the second map is that, it is not symplectic, but it approaches the second and the first original map when we increase the order of the square matrix, and the perturbation becomes smaller and smaller.

The central topic of this work is to study the differences between the second map (hence the first map, obtained from tracking simulation) and the four'th map, an idealized integral system,  using the characteristic functions $\Delta, \text{Im}(\phi)$, obtained from the square matrix analysis.

\subsection{Comparison with Normal Form}
There is apparently a similarity between the square matrix method and the normal form.	However the similarity is only superficial. The differences between the two methods are very obvious, we name a few here:

1.	We emphasize that we only need one straight forward step of Jordan decomposition to derive the high order result while the normal form requires order by order iteration, thus making the procedure very complicated.

2.	The tune expression is a rational function of the variables for the square matrix method while the normal form for the tune is a polynomial, so it is obviously very different.

3.	We obtain the expression for the tune fluctuation and amplitude fluctuation, which is essential in its application to predict the dynamical aperture in a practical way while the normal form does not have such kind of expression because the frequency variation is intrinsically associated with the non-integrability of the system.

The essential feature of the square matrix method is that it is easy to reach very high order with high precision. Hence a detailed comparison requires works on both square matrix and normal form. Obviously this requires a significant work.
Clearly the best way is to compare with the exact answer from particle tracking. The purpose of this paper is to provide examples of this comparison as outlined in the following.

\subsection{Outline}
First, in Section 2, as an illustration for the basic principle, we give one simple example of the solution of a nonlinear differential equation. We apply the square matrix method given above to solve this equation to lowest nonlinear order of 3, and compare the method with the well known result given in the text book by Landau and Lifshitz \cite{landau_1969}.

In Section 3 we apply the square matrix method to the one turn map of storage ring lattice to compare the square matrix analysis with simulation.
In Section 4 we present one example of the application, the manipulation of phase space trajectory.

The applications of the square matrix analysis in Section 2 to 4 are all based on the variable transformation from $z$ to the action-angle variable $w_0$ in Eq.(1.11), obtained from the left eigenvectors $U$ in Eq.(1.7). This matrix $U$ is obtained by Jordan decomposition of the mapping matrix $M$ in Eq.(1.3). For high order, the dimension of $M$ is large. Hence the applications described in Section 3 and 4 depend on the efficient Jordan decomposition of very large dimension matrix $M$ with high stability and precision.

In Section 5,6, we explain how to achieve the efficient Jordan decomposition.
Since the details of this are often technical and involve some abstract mathematics, we only present the outline and leave the details in the appendixes.

In Section 7 we analyze the multi-turn behavior of the action-angle variable $w_0$, and show that the  phase advance has a term $ \Delta n^2 /2$, as given in Eq.(1.20). This section is an important section because it gives a more clear physical meaning to the quantities such as $w_0, \phi$, and $\Delta$, used in the applications in Section 3 and 4. The fact that we present these after Section 5 and 6, which involve more mathematics, is because it requires some of the basic concepts such as the chain structure of the Jordan decomposition given in Section 5,6.

In Section 8 we show how to improve the stability and precision while ensuring the uniqueness of the Jordan decomposition, which is crucial for the square matrix method. Finally, Section 9 gives a summary.
\section{One example: The solution of a nonlinear differential equation}
\label{sec:section2}
As an illustration of the square matrix method, we consider the differential equation given in the text book by Landau and Lifshitz \cite{landau_1969}:
\begin{equation}
   \label{eq:Eq.2.1}
   \begin{split}
  \ddot{x}+\omega_0^2 x= - \alpha x^2 - \beta x^3
  \end{split}
\end{equation}
In order to apply the square matrix method, we first transform to normalized coordinates $z \equiv \bar{x}- i \bar{p}, z^\ast \equiv \bar{x}+ i \bar{p}$, where
$\bar{x} \equiv \omega_0 x, \bar{p} \equiv \dot{x}$. The differential equation becomes
\begin{equation}
   \label{eq:Eq.2.2}
   \begin{split}
  \dot{z}&=\hspace{3mm}  i \omega_0 z + i a (z+z^\ast)^2 +i b  (z+z^\ast)^3 \\
    \dot{z^\ast}&= -i \omega_0 z^\ast - i a (z+z^\ast)^2 -i b  (z+z^\ast)^3 ,\hspace{2mm} \text{with}\\
a &\equiv \alpha/(4 \omega^2), b \equiv \beta /(8 \omega^3)
  \end{split}
\end{equation}
  Following the steps in Section 1A, we can write a square matrix equation $\dot{Z}= M Z$. We remark that this equation is different from the Eq(1.3), where the left hand side of the equation is the new column $Z$ after one turn, while here we have the time derivative of the column $\dot{Z}$ because in this example, we discuss a differential equation rather than a one turn map. But the basic principle for solution is the same.  For example, the third row in $\dot{Z}= M Z$ is
\begin{equation}
   \label{eq:Eq.2.3}
   \begin{split}
  \frac {d (z^2)}{dt} &=2 z \dot{z}=2  i \omega_0 z^2 + 2i az (z+z^\ast)^2 +2i b  z(z+z^\ast)^3 \\
  \end{split}
\end{equation}
Expand the polynomial and continue to derive the derivative of the monomials in $Z$, truncated at 3rd order, the last row is
\begin{equation}
   \label{eq:Eq.2.4}
  \frac {d (z^{\ast3})}{dt} =3 z^\ast2 \dot{z^\ast}  = -3  i \omega_0 z^\ast
\end{equation}
Combining these results, we find $\dot{Z}= M Z$, with
{\footnotesize
\begin{equation}
   \label{eq:Eq.2.5}
M=
\begin{bmatrix}
  i\omega_0 & 0 & ia & 2ia & ia & ib & 3ib & 3ib& ib \\
0 &   -i\omega_0 & -ia & -2ia & -ia & -ib & -3ib & -3ib& -ib \\
0 &   0 &2 i\omega_0 & 0 &0 & 2ia & 4ia & 2ib& 0 \\
0 &  0 &0&0&0 & -ia & -ia & ia& ia \\
0 &  0 &0&0&-2i\omega_0 &0& -2ia &-4 ia& -2ia \\
0 &  0 &0&0&0 & 3i\omega_0 & 0 & 0& 0 \\
0 &  0 &0&0&0 & 0 & i\omega_0 & 0& 0 \\
0 &  0 &0&0&0 & 0 &0 & -i\omega_0& 0 \\
0 &  0 &0&0&0 & 0& 0 & 0& -3i\omega_0 \\
 \end{bmatrix}
\end{equation}
}
The $9 \times 9$ matrix $M$ has 9 left eigenvectors corresponding to each of the 9 diagonal elements of $M$. Two of them have eigenvalue $i \omega_0$.
We arrange these two eigenvectors as a column of two rows $U=
  \begin{bmatrix}
  u_0\\u_1
   \end{bmatrix}=$
{\scriptsize
\begin{equation}
   \label{eq:Eq.2.6}
\begin{split}
&\begin{bmatrix}
1 & 0 &-\frac{a}{\omega_0} &  \frac{2a}{\omega_0}  &  \frac{a}{3\omega_0}  & \frac{4a^2-b\omega_0}{2\omega_0^2} & 0& \frac{-4a^2+9b\omega_0}{6\omega_0^2} &\frac{4a^2+3 b\omega_0}{12\omega_0^2}  \\
   0 & 0 &0&  0 & 0 & 0&3ib- \frac{20ia^2}{3\omega_0} &0&0  \\
   \end{bmatrix}
\end{split}
\end{equation}
}
One can check indeed that similar to the left eigenvector equation Eq.(1.7), $U$ satisfies the left eigenvector equation $UM=(i \omega_0 I+\tau)U$, with I the $2 \times 2$ identity matrix, and $\tau=
\begin{bmatrix}
0&1\\0&0
   \end{bmatrix}$, the Jordan matrix.
Now, as in Eq.(1.11), we define $W=UZ = \begin{bmatrix}
  u_0Z\\u_1Z
   \end{bmatrix} \equiv \begin{bmatrix}
  w_0\\w_1
   \end{bmatrix}$. Then
\begin{equation}
   \label{eq:Eq.2.7}
\dot{W}=U \dot{Z}=UMZ=(i \omega_0 I+\tau)UZ=(i \omega_0 I+\tau)W
\end{equation}
The solution of this matrix equation is
\begin{equation}
   \label{eq:Eq.2.8}
\begin{split}
W=e^{(i\omega_0+\tau)t}W_0
\end{split}
\end{equation}
where $W_0$ is the initial value of $W$. Since this solution is based on truncated power series of $z,z^\ast$, it is an approximation, valid for a finite time interval. Within this approximation, the matrix $\tau$ is to be replaced by a number $i \phi$ so that $\phi$ represents a frequency shift. That is, $W_0$ should be an eigenvector of the matrix $\tau$ with eigenvalue $i \phi$, and every row of $W_0$ should be an action-angle variable, up to the order of the expansion.

The two rows of the matrix equation Eq.(2.7) are
\begin{equation}
   \label{eq:Eq.2.9}
\begin{split}
 &\dot{w_0}= i \omega_0 w_0 + w_1=(i \omega_0 +\frac{ w_1}{w_0})w_0 \equiv i (\omega_0 +\phi )w_0 \\
 &\dot{w_1}= i \omega_0 w_1   \\
\end{split}
\end{equation}
From the top row, we identify $\phi=-i \frac{w_1}{w_0}$ as the frequency shift. Using Eq.(2.6), we have
{\footnotesize
\begin{equation}
   \label{eq:Eq.2.10}
\begin{split}
&w_0 \equiv u_0Z=z -\frac{a}{\omega_0} z^2+  \frac{2a}{\omega_0} zz^\ast +  \frac{a}{3\omega_0} z^{\ast2}\\
&+ \frac{4a^2-b\omega_0}{2\omega_0^2} z^3  + \frac{-4a^2+9b\omega_0}{6\omega_0^2} zz^{\ast2} +   \frac{4a^2+3 b\omega_0}{12\omega_0^2} z^{\ast3} \\
& w_1 \equiv u_1Z=(3ib- \frac{20ia^2}{3\omega_0} )z^2z^\ast  \\
\end{split}
\end{equation}
}
To the lowest order approximation, in deriving the frequency shift, we only keep the first term $z$ in $w_0$ of Eq.(2.10), substitute the expression of $a,b$ in eq.(2.2) and the relation between $z$ and $x$, take $\dot{x}=0$, get the amplitude dependent frequency shift by Landau \cite{landau_1969}:
{\small
\begin{equation}
   \label{eq:Eq.2.11}
\begin{split}
\Delta \omega=\phi \cong (3b- \frac{20a^2}{3\omega_0} )zz^\ast = (\frac{3 \beta}{8 \omega_0}- \frac{5 \alpha^2}{12 \omega_0^3} ) x^2\\
\end{split}
\end{equation}
}
\begin{figure}[htb]
  \centering
  \includegraphics[width=\columnwidth]{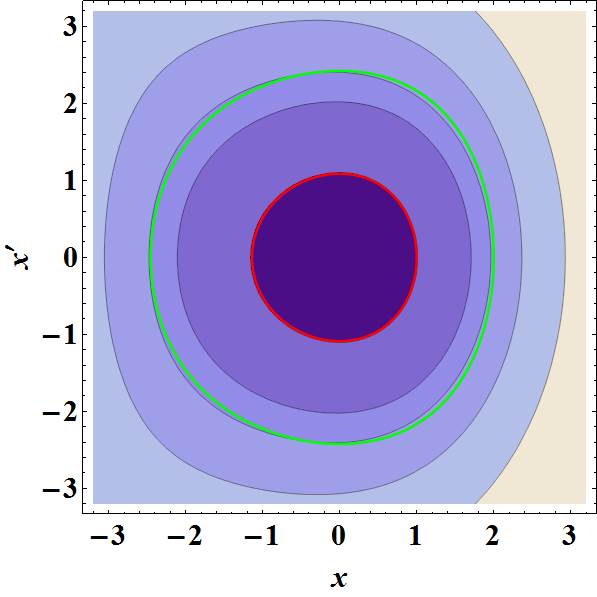}
  \caption{\label{fig:fig01} Contour plot of $|w_0|$ in $x,\dot{x}$ plane, compare with the trajectory of the solutions of Eq.(2.1) for initial x=1 (red),x=2 (green), $\omega_0=1,\alpha=0.2,\beta=0.1$.}
\end{figure}\\
As action-angle variable, $|w_0|,|w_1|, |\phi|=|w_1/w_0|$ are invariant. In Fig.1 we plot the contours of $|w_0|$ in the $x,\dot{x}$ plane for a case where $\omega_0=1,\alpha=0.2, \beta=0.1$, in comparison with the solutions of Eq.(2.1). It is clear that even when initial amplitude is as large as x=2, the trajectory of the solution, even though significantly deviates from a circle, is still in good agreement with the constant contour of $|w_0|$, which corresponds to a circle in $w_0$ complex plane. Hence the terms in $w_0$ with order higher than 1, provide correct information about the nonlinear distortion of the trajectory.

When we examine Eq.(2.10), we observe that $w_1$ has only a 3rd order term, hence this term is the approximation of the lowest order for $w_1$. Actually, up to 3rd order the term $|z(zz^\ast)|=|z|^3$ represents a circle in the complex plane of $z$, the higher order terms needed to represent the nonlinear distortion and the frequency shift are neglected because the truncation at 3rd order for $w_1$.

To calculate the frequency shift more accurately, we take the definition of $w_0$ as a variable transformation from $z,z^\ast$ to $w_0,w_0^\ast$, then the top row of Eq.(2.9) can be taken as a differential equation for $w_0$. To clarify this point, we remark that $\phi(z)=-i w_1(z)/w_0(z)$ is a function of z in Eq.(2.9), and we take $z(w_0)$ as the inverse function of $w_0(z)$ given by Eq.(2.10), then we consider $\phi$ in Eq.(2.9) as an implicit function of $w_0$. Hence even $\phi(w_0)=\phi(z(w_0))$ is an invariant approximately, it evolves with time and has small deviation from constant.

We take $w_0=r e^{i \theta }$. Based on the discussion following Eq.(2.11), $r$ is nearly a constant, up to 3rd order. Hence $\phi$ is a function of $\theta$. When substituted into Eq.(2.9), we get
{\small
\begin{equation}
   \label{eq:Eq.2.12}
\begin{split}
\frac{d \theta}{dt} =\omega_0 + \phi(\theta) \hspace{2mm}
\end{split}
\end{equation}
}
We can solve Eq.(2.12) along a circle as a function of $\theta$ in the $w_0$ plane and find the period $T$:
{\small
\begin{equation}
   \label{eq:Eq.2.13}
\begin{split}
T=\int_0^T dt= \int_0^{2 \pi} \frac {d \theta} {\omega_0 + \phi( \theta)}
\end{split}
\end{equation}
}
From the period $T$ we calculate the frequency shift as $\Delta \omega=2  \pi/ T-\omega_0= 1 / < \frac {d t}{d \theta}>-\omega_0$, where $< \frac {d t}{d \theta}>$ is an average over a period. In Fig.2 we plot the frequency shift obtained from this result, and compare with the Landau formula Eq.(2.10), and the frequency calculated from the direct numerical solution of Eq.(2.1), for the case of $\omega_0=1, \alpha=0.2, \beta=0.1$, showing that indeed this higher order calculation provides a much better agreement at larger amplitude of x=2 than the Landau formula.
\begin{figure}[htb]
  \centering
  \includegraphics[width=\columnwidth]{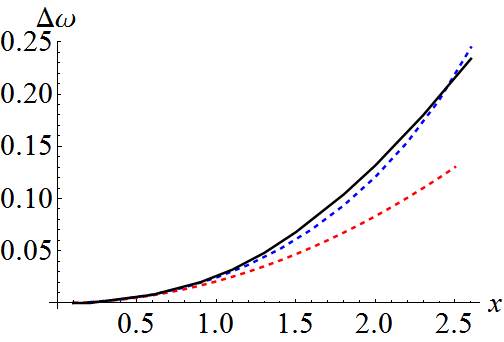}
  \caption{\label{fig:fig02} The frequency shift, as function of initial x, calculated from numerical integration of Eq.(2.1) (black solid) compared with Landau formula Eq.(2.11) (red dash), and 3rd order square matrix calculation (blue dash), $\omega_0=1,\alpha=0.2,\beta=0.1$.}
\end{figure}

It is clear that at 3rd order the agreement with exact solution is excellent. We remark that in the square matrix method, the calculation is carried out in one step to 3rd order without iteration steps. For first and second order, there is only one diagonal element in $M$ equal to $i \omega_0$ and the eigenspace for $i \omega_0$ has only dimension 1, hence we directly start the calculation at 3rd order. For the traditional canonical perturbation method, the calculation is carried out order by order from low to high by complicated iteration procedure. Hence it seems these two methods are not completely equivalent. The advantage of the square matrix approach is that it only need one step to reach high order, and its procedure is simple and straight forward. In the next section we apply this method to the analysis of  the one turn map of the storage ring lattices, and compare with simulation.

\section{Comparison with Simulation }
\label{sec:section3}
\subsection{Summary about application of the theory}
When we study the nonlinear dynamic equations such as Hill equations, in order to have the square matrix  to be triangular, we always first convert the variables x, p into normalized Courant-Snyder variables $\bar{x},\bar{p}$  using the betatron amplitude matrix $B^{-1}$ (see, e.g., S.Y. Lee, p.49 \cite{sylee_2012}). Then they are converted to scaled variables using a scaling parameter s, as will be given in Section 8:
\begin{equation}
  \label{eq:Eq.3.1}
z=\bar{z}/s \equiv (\bar{x}-i\bar{p})/s, z^\ast=\bar{z^\ast}/s \equiv (\bar{x}+i\bar{p})/s,
\end{equation}
where
\begin{equation}
   \label{eq:Eq.3.2}
   \begin{split}
  \begin{bmatrix}
   \bar{z} \\ \bar{z^\ast}
   \end{bmatrix}&=K^{-1}
  \begin{bmatrix}
 \bar{x} \\ \bar{p}
   \end{bmatrix} = K^{-1} B^{-1}
\begin{bmatrix}
  x \\  p
 \end{bmatrix},\\B^{-1}&=
   \begin{bmatrix}
\frac{1}{\sqrt{\beta}} & 0 \\
  \frac{\alpha}{\sqrt{\beta}} & \sqrt{\beta}
   \end{bmatrix}, K^{-1}=
      \begin{bmatrix}
1 & -i \\
  1 & i
   \end{bmatrix}
   \end{split}
\end{equation}
For 4 variables such as $x,p_x,y,p_y$ , it is similar, with both $B^{-1}$ and $K^{-1}$ replaced by 4$\times$4 matrixes. But we must first carry out linear decoupling between x and y. As we shall point out in Section 8, we select the scaling factor $s$ for the variable  to minimize the range of the coefficients in the square matrix, in order to ensure the stability and precision of the Jordan decomposition.

During our calculation we use the Taylor expansion provided by the well-known program of TPSA (truncated power series algorithm)\cite{berz_1989} as our starting point of the one turn map corresponding to Eq.\,\eqref{eq:Eq.1.1}. We construct the square matrix $M$ up to a certain order. Then we apply Jordan decomposition for the invariant subspaces to obtain the transformation matrix $U$ for the specific eigenvalues $e^{i\mu_x}, e^{i\mu_y}$. As described in Section 1B, we use the first row of $U$ to define the variable transformation $w_0=u_0Z$. This serves as an accurate action-angle variable (for 4 variables case we have $w_{0x}=u_{0x}Z, w_{0y}=u_{0y}Z $). Then the 2nd and 3rd row of $U$ are used to calculate the functions $\phi$ and $\Delta$  using Eq.\,\eqref{eq:Eq.1.17} and Eq.\,\eqref{eq:Eq.1.19}. We remark here that these are no longer polynomials, they are rational functions.
\begin{figure}[htb]
  \centering
  \includegraphics[width=\columnwidth]{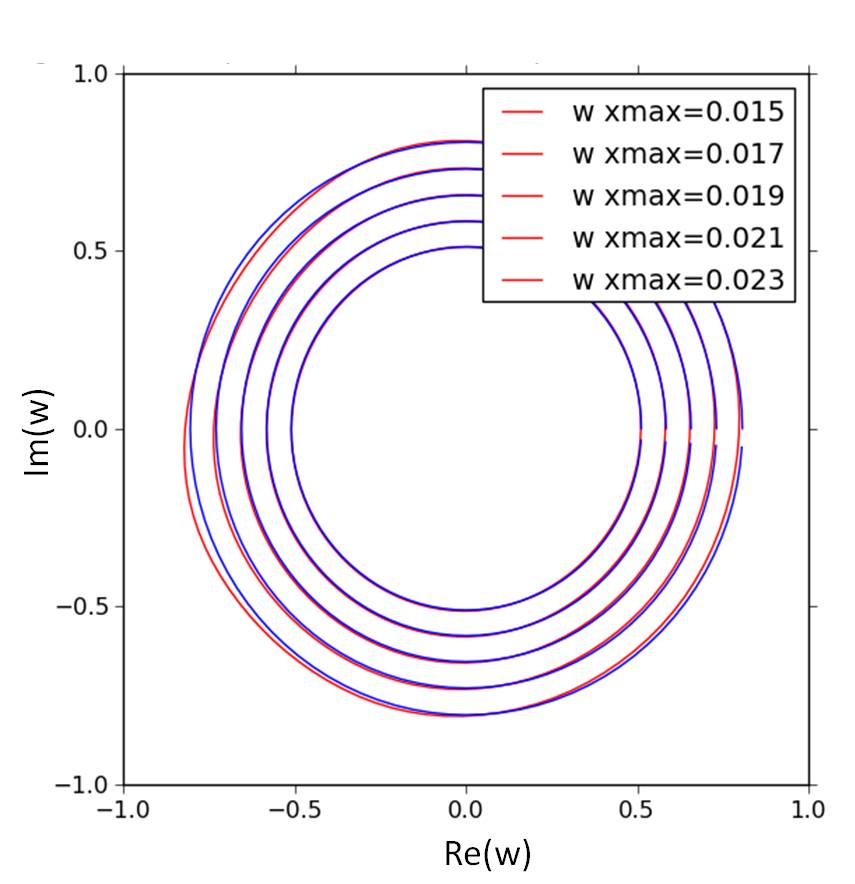}
  \caption{\label{fig:fig3} Curves (blue) with constant $|w_x |$ at 7th order in w phase space compared with result of tracking one turn (red) }
\end{figure}

\begin{figure}[htb]
  \centering
  \includegraphics[width=\columnwidth]{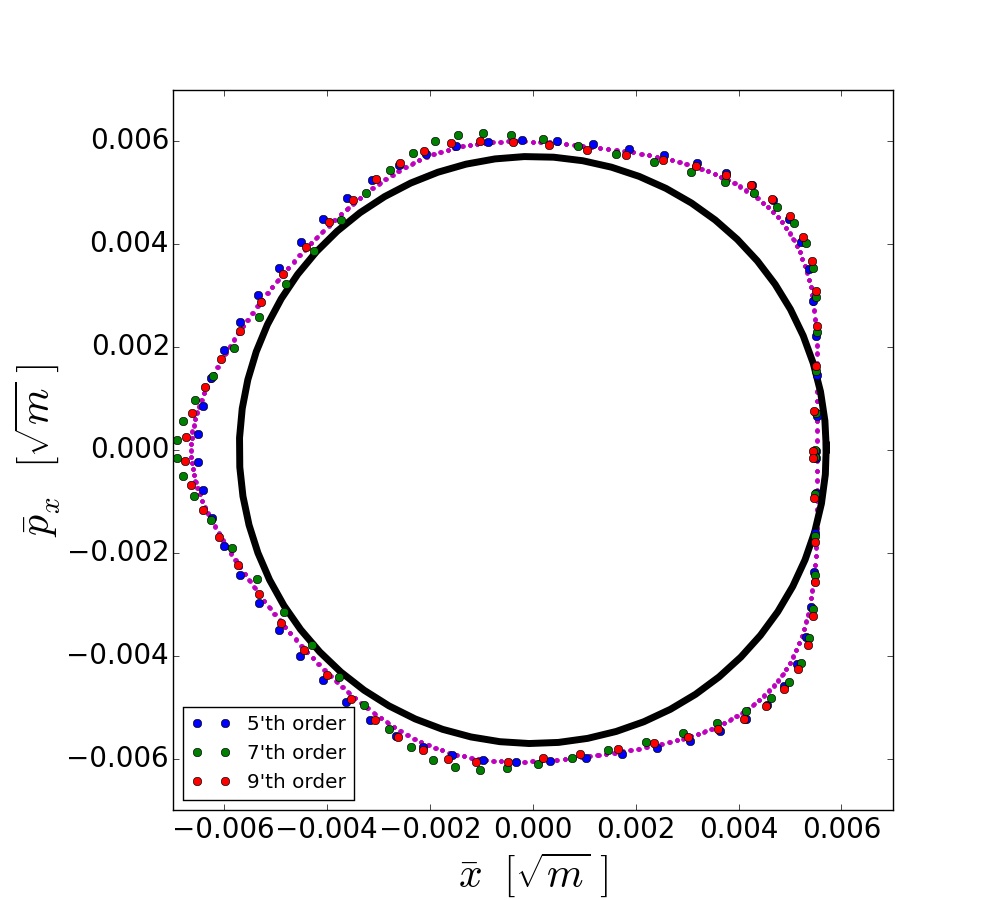}
  \caption{\label{fig:fig4} Curves with constant $|w_x|$ at 5,7,9th order (blue, green, red) in   phase space compared with result of tracking 512 turns and the circle passing through initial point (black)}
\end{figure}
\subsection{Action-angle variable and Stability near Invariant Tori Border}
\begin{figure}[htb]
  \centering
  \includegraphics[width=\columnwidth]{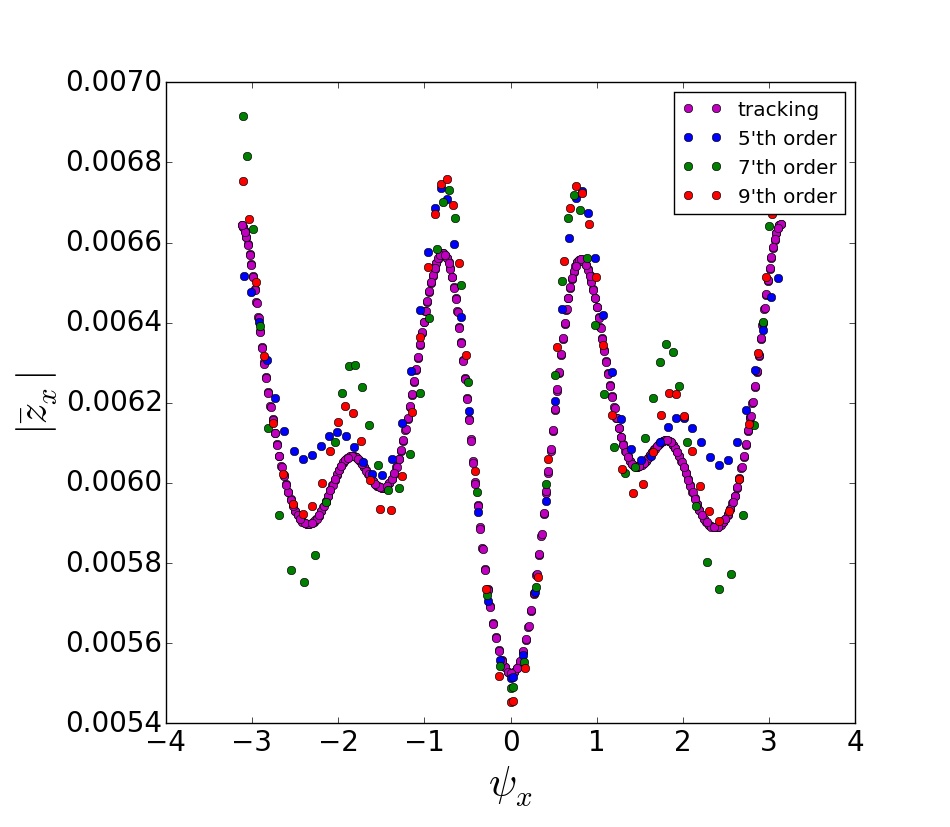}
  \caption{\label{fig:fig5} Detail of curves in Fig.4 with constant $|w_x|$ at 5,7,9th order (blue, green, red) in $\bar{z}_x$ phase space compared with result of tracking 512 turns(magenta)}
\end{figure}
\begin{figure}[htb]
  \centering
   \includegraphics[width=\columnwidth]{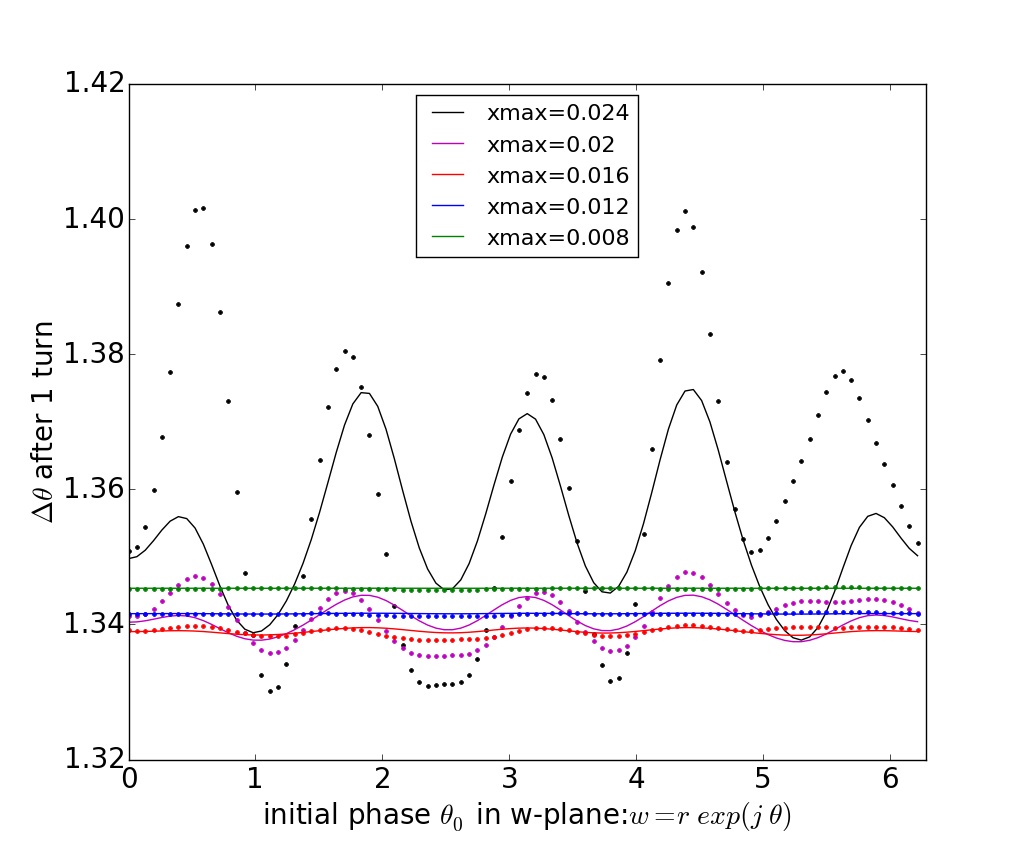}
  \caption{\label{fig:fig6} The $\Delta \theta$ after one turn calculated from tracking (dotted line), and from $\text{Re} \phi+\text{Im} \Delta /2$ (solid line) for various initial x=8,12,16,20,24mm}
\end{figure}
\begin{figure}[htb]
  \centering
 \includegraphics[width=\columnwidth]{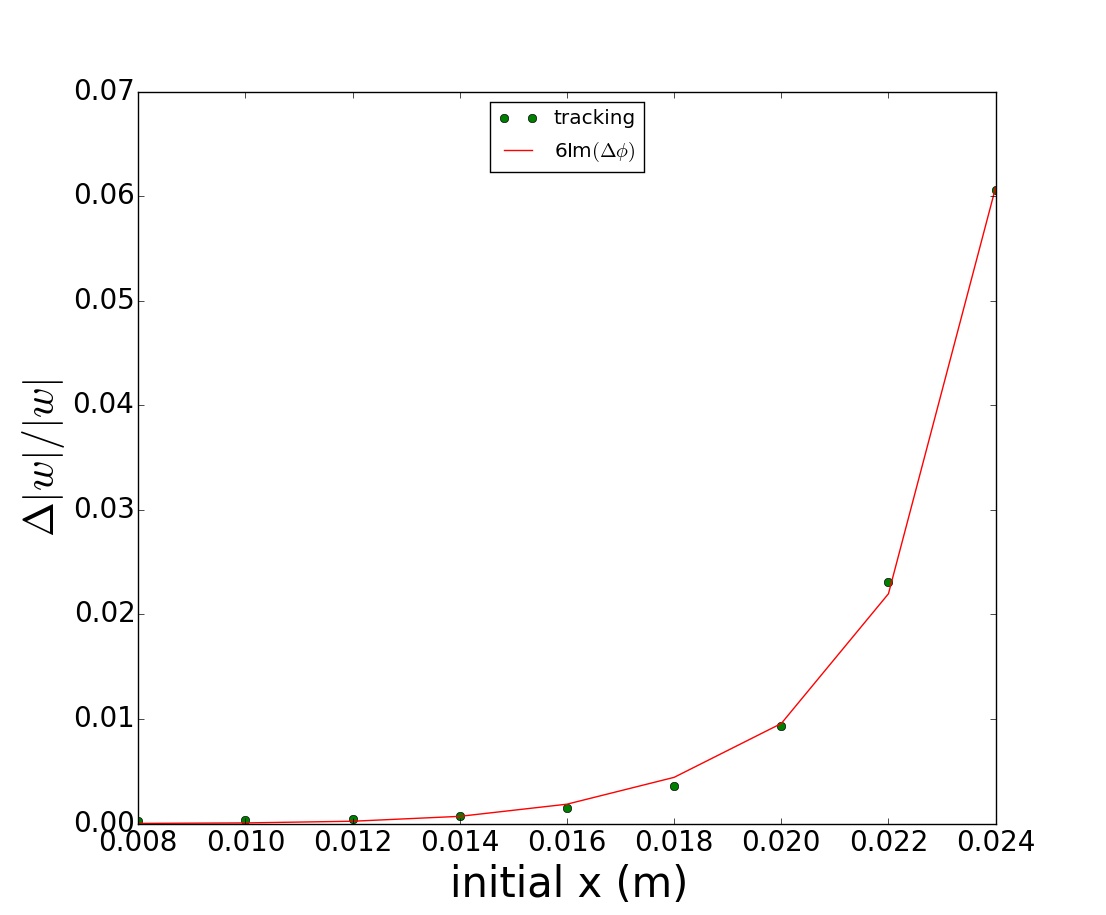}
  \caption{\label{fig:fig7}  plot the difference of maximum and minimum of $|\Delta w /w|$ as function of initial x. Tracking (green) is compared with prediction by  (red) multiplied by a constant to obtain agreement with  the tracking}
\end{figure}
The equations
\begin{equation}
\label{eq:Eq.3.3}
w_x \equiv w_{0x}(z_x,z_y), w_y \equiv w_{0y}(z_x,z_y)
  \end{equation}
define the set of variables $w_x,w_y,w_x^\ast,w_y^\ast$ as functions of $z_x,z_y,z_x^\ast,z_y^\ast$ . Notice that because $w_0$, representing the first row of $W$, is used very often, we simply use $w$ to represent it, and from now on often we use $w_{x0}$ to represent its initial value when we specify it.  When the “coherence” conditions Eq.\,\eqref{eq:Eq.1.17}, Eq.\,\eqref{eq:Eq.1.18} and Eq.\,\eqref{eq:Eq.1.19} are satisfied, these two equations provide a transformation from $z_x,z_y $ to the new variables as a set of accurate action-angle variables.

The inverse function $z_x(w_x,w_y), z_y(w_x,w_y)$ of Eq.(3.3) is very useful because a set of constant $r_x \equiv |w_x|$,$r_y \equiv |w_y|$ describes the motion of the particles. Appendix D shows that the inverse function can be calculated using the inverse of a upper-triangular matrix in a simple way.
	
In Fig.3, for a lattice "EPU"of NSLSII ("National Synchrotron Light Source II") with all insertion devices included, for the points around a circle in w-plane (i.e., a blue circle with r=constant), we use the inverse function of $w_x(z_{x0})$ to find a set of initial $x_0,p_{x0}$  ($y_0,p_{y0}$ are set zero), then after tracking these particles for one turn, calculate $w_x(z_{x})$ and plot the red curves. We can see that when x approaches the dynamic aperture at x=25mm, $w_x$ (red) gradually deviates from the circles (blue).
\begin{figure*}[htb]
  \centering
 \includegraphics[width=0.48\textwidth]{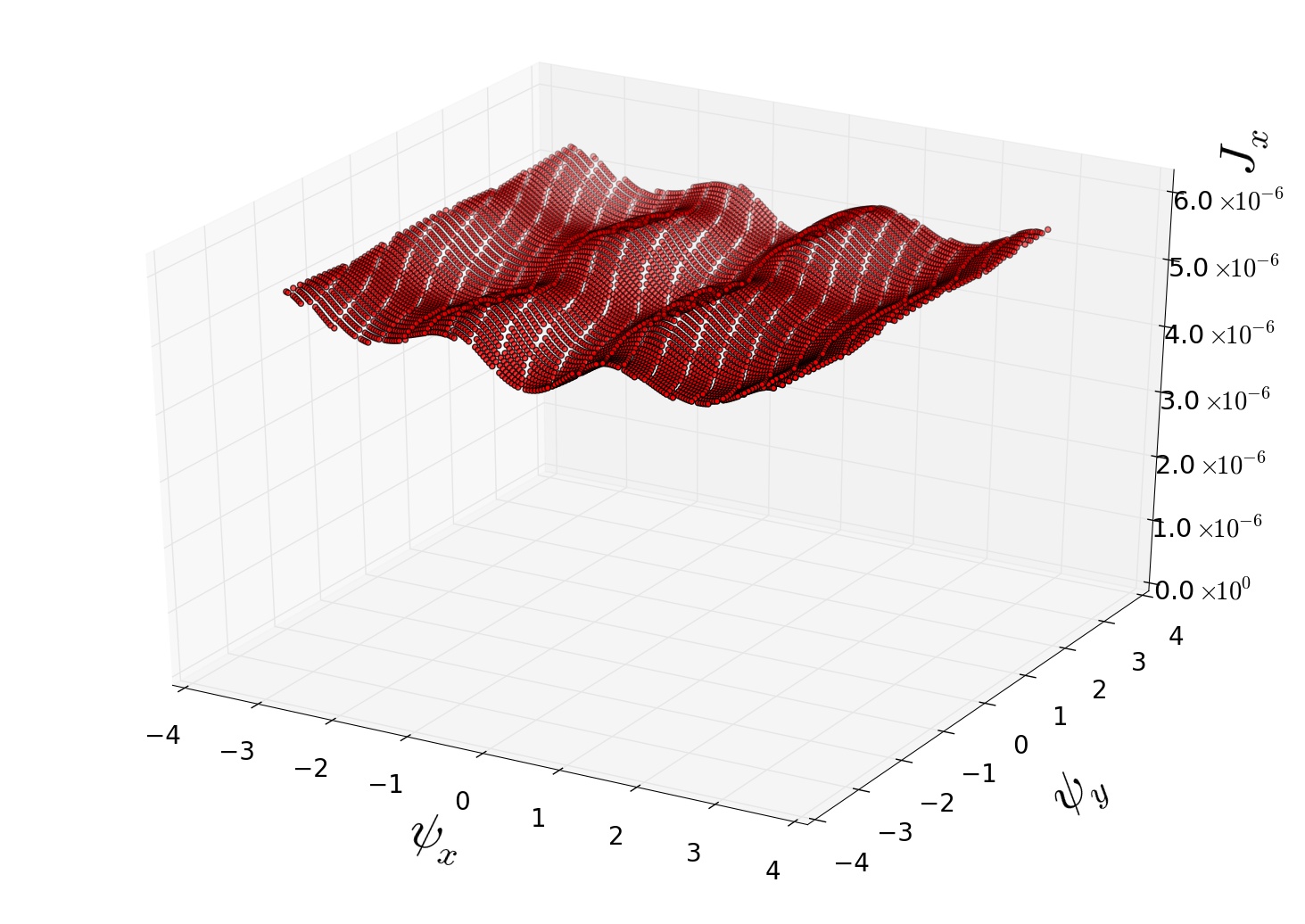}
  \includegraphics[width=0.48\textwidth]{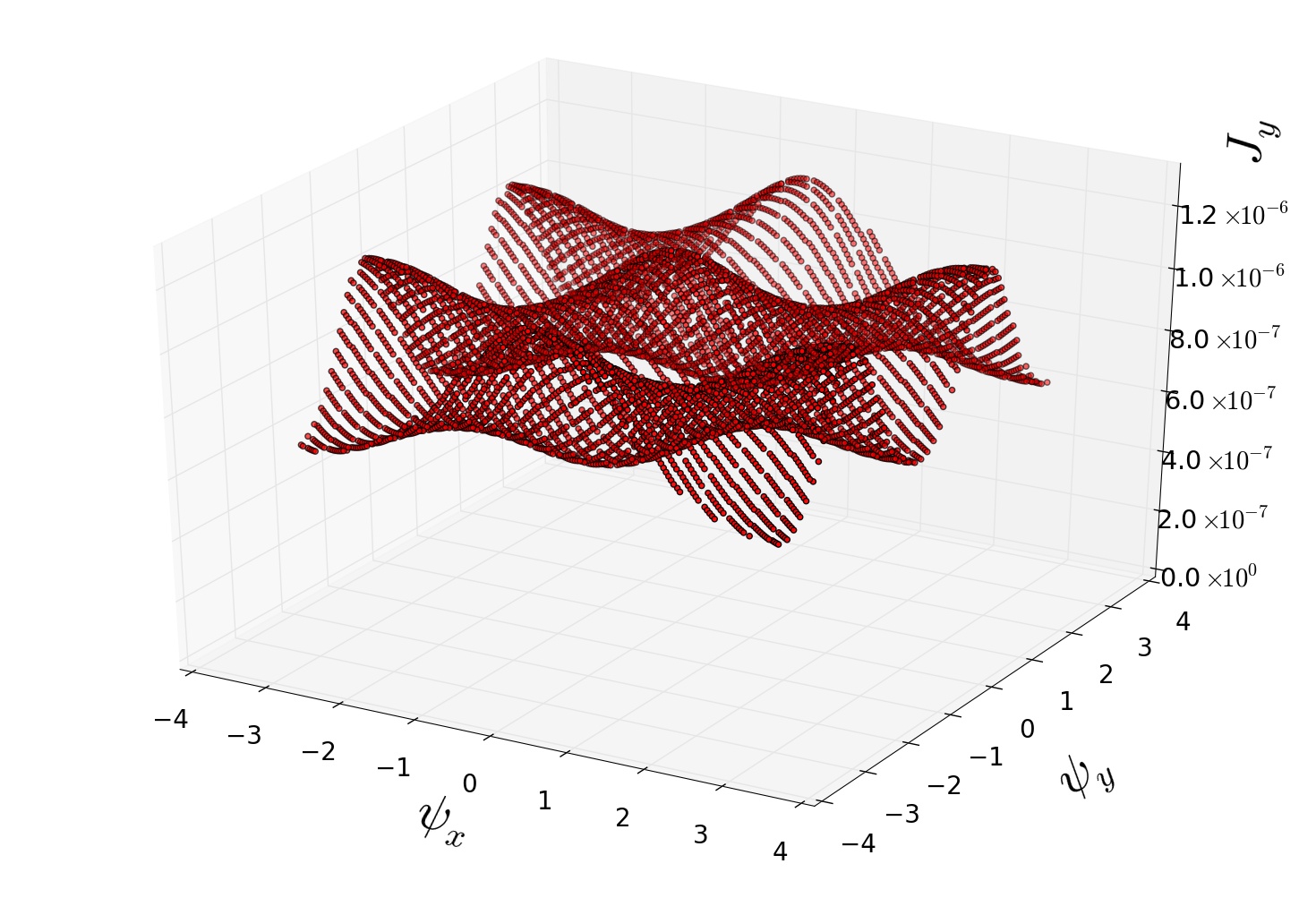}\\
   \includegraphics[width=0.48\textwidth]{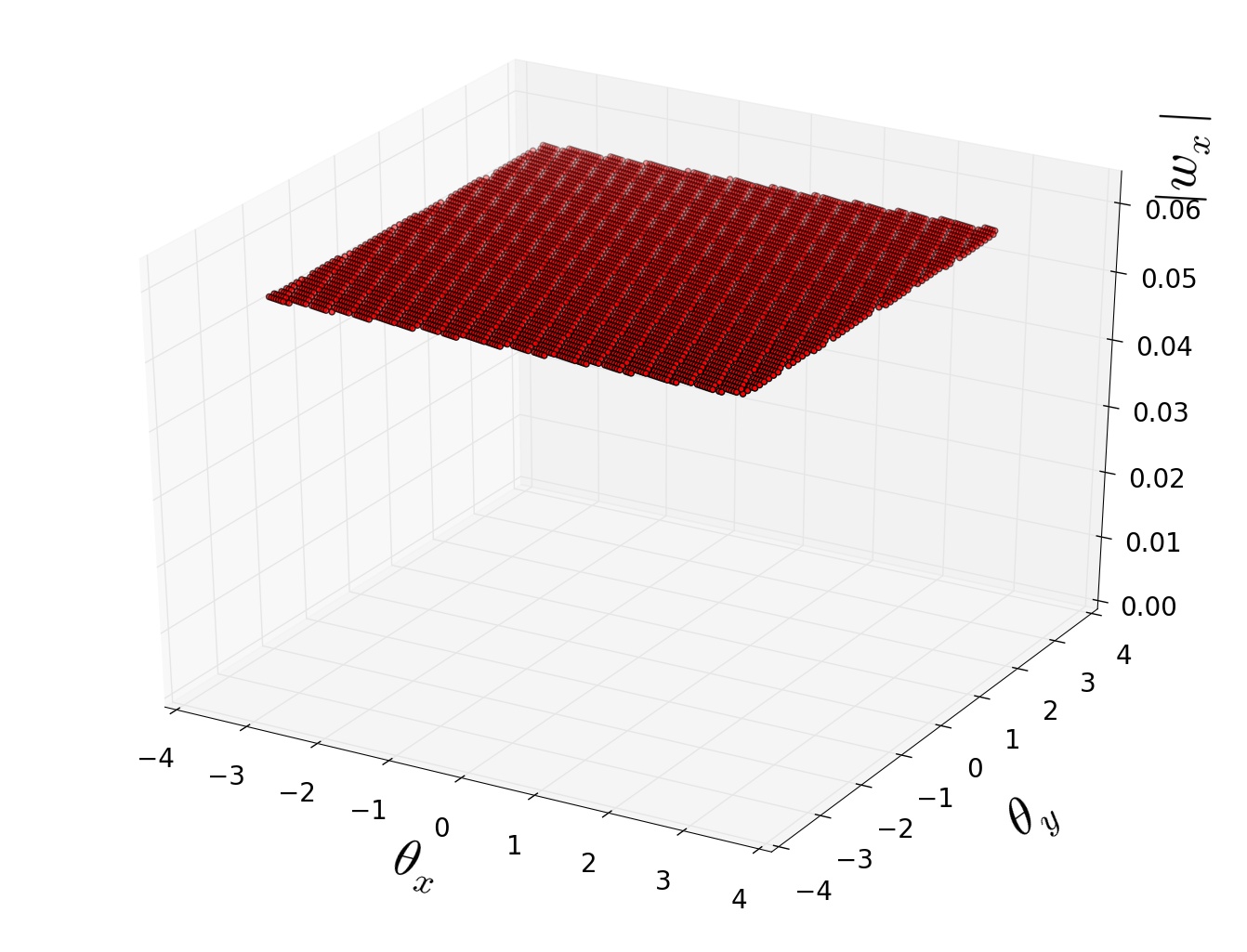}
    \includegraphics[width=0.48\textwidth]{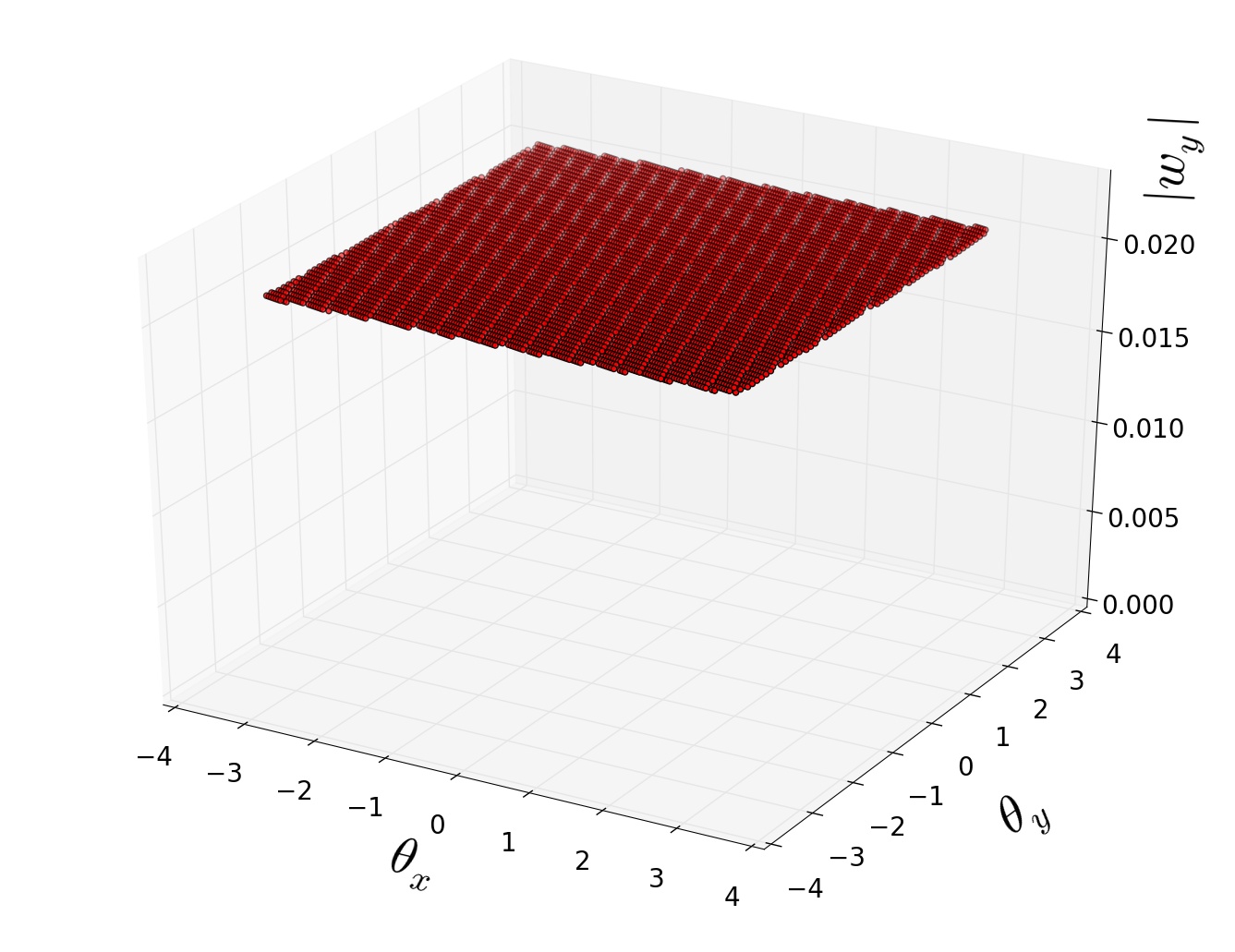}
  \caption{\label{fig:fig9}  Poincare surface of section using Courant-Snyder normalized variables $\bar{z}_x,\bar{z}_y$ (top row) compared with new variables $w_{x},w_{y}$ (bottom row). Left two plots are for x, right are for y. Initial values of $|w_{x0}|$,$|w_{y0}|$ for all points are same as a point generated from x=10mm, y=2mm}
\end{figure*}

In Fig.4, for another one of the lattices of NSLSII, we plot on $\bar{z}_x$ -plane a circle which passes through the point corresponding to initial $x_0=25mm,p_{x0}=0$, and plot the tracking result over 512 turns as the magenta curve. Let $w_x=r e^{i \theta}$ and initial $r_{0} \equiv |w_{x0}|$ we also plot all the points calculated according to a constant $|w_x|=r_{0}$. The calculation is up to 5 (blue), 7(green), and 9(red) order, respectively.

It is clear the agreement is excellent. To see the errors of different order, we plot the same set of data in Fig.5 with fine details, showing $|\overline{z}_x|$ as function of angle $\psi_x$ (the phase angle of $\bar{z}_x=J_x e^{i\psi_x}$). We can see that as the order for the constant $|w_x|$  increases, the agreement with the tracking result (magenta curve) converges slowly, with the 9’th order (red) more close to the tracking.

In Fig 6 we show the  the phase advance $\Delta \theta$ after one turn as functions of initial phase $\theta_0$ for various initial $r_{0}$, which corresponds to different initial $x_0$. The dotted curves are from tracking, solid curves are calculated from $\text{Re}\phi+\text{Im}\Delta /2$. It is clear from Fig.6 that as r increases, $\Delta \theta$ have larger and larger variation. For large amplitude x=22mm, we can see that the two curves do not agree with each other, even though the trend of increased variation is obvious. The theoretical prediction on $\Delta \theta$ is given by the variation of $ \text{Re}\phi+ \text{Im}\Delta /2$ as function of $\theta$. These variations are an indication of a deviation from “coherence”, i.e., a violation of the condition given by Eq.\,\eqref{eq:Eq.1.17}. Thus when this condition is violated, the calculation $\phi$ lost precision, hence they deviate from the tracking results. However, even though the fluctuation of $\phi$ does not accurately predict the deviation, it still provides information about the deviation from coherence.

Similarly, the fluctuation of $r$ after one turn starting from a circle of constant $r_0$ also provides information about the deviation from coherence, approximately agrees with the prediction of $\text{Im} \phi$. In Fig. 7, we plot the peak to peak deviation of $|\Delta w /w|$ as function of initial x. We see that $ \text{Im} (\Delta \phi)$  does not give accurate $|\Delta w /w|$, so that we need to multiply $ \text{Im} (\Delta \phi)$ by a factor 6 to obtain agreement with $|\Delta w /w|$ found from tracking.  This is because $ \text{Im} (\Delta \phi) \neq 0$ itself implies the theory lost its precision. But it is seen from this plot that the deviation from coherence is predicted by $ \text{Im} (\Delta \phi)$ correctly, and it does serve as an index for the proximity to the destruction of invariant tori.

Next, we check the cases of 4 variables $x,p_x,y,p_y$. We study a lattice named nsls2sr\_supercell\_ch77. In Fig.8, the top row is the Poincare surface of section \cite{lieb_1983, ruth_1986} expressed by the Courant-Snyder variable $\bar{z}_x= \bar{x}-i\bar{p}_x $ and $\bar{z}_y= \bar{y}-i\bar{p}_y $. The horizontal axes are their phase angles $\psi_x,\psi_y$ respectively. The vertical axes are their amplitude $|\bar{z}_x|$,  $|\bar{z}_y|$. The left plot is for the amplitude $|\bar{z}_x|$, the right one is for $|\bar{z}_y|$. For the case of initial x=10mm, y=2mm, we track the particle for 512 turns. Every point on these plots is obtained from the coordinates for a specific turn. For the same set of data, when we convert $\bar{z}_x$, $\bar{z}_y$ to $w_x,w_y $ and plot the Poincare sections for the corresponding variable $\theta_x,\theta_y$  (the phase angle of  $w_x=r_x e^{i\theta_x}, w_y=r_y e^{i\theta_y}$ ) as the transverse axes, and  $ r_x =| w_x |, r_y=| w_y|$  as the vertical axes, we obtain the bottom row of Fig.8. Clearly, these new variables now move on two separate flat planes in the two Poincare sections, representing two independent rotations. Thus the transformation to new variables $w_x,w_y $ reduces the very complicated motion expressed by $z_x,z_y $ to two very simple uniform independent rotations.

\subsection{Amplitude Dependent Tune shift and Tune footprint}
In Fig.9 we plot tune $\nu_x$ as function of initial x for various y position (initial $p_x =p_y =0$), compare tune from tracking result (green) with tune calculated from $\mu+\text{Re}\phi$  (red) using Eq.\,\eqref{eq:Eq.1.17}. There is an excellent agreement up to near the dynamic aperture. We see that at y=6mm and y=-6mm, when the x passes x=-1mm, there is a resonance. We can see the green curve (tracking) has a discontinuity, and the red curve (the square matrix derived tune) also has a jump. Near this point $\nu_x \approx 2 \nu_y$, we have two frequencies dominate the spectrum of x motion: $\nu_x , 2 \nu_y$. The single frequency condition is no longer valid. Hence the “coherence” condition Eq.\,\eqref{eq:Eq.1.17} is violated. Even though the
red curves seem to exaggerate the discontinuity, it does show the resonance clearly. This suggests that the square matrix analysis may provide more detailed understanding about resonances. However, we will not discuss about resonances in general in this paper.
\begin{figure}[htb]
  \centering
\includegraphics[width=\columnwidth]{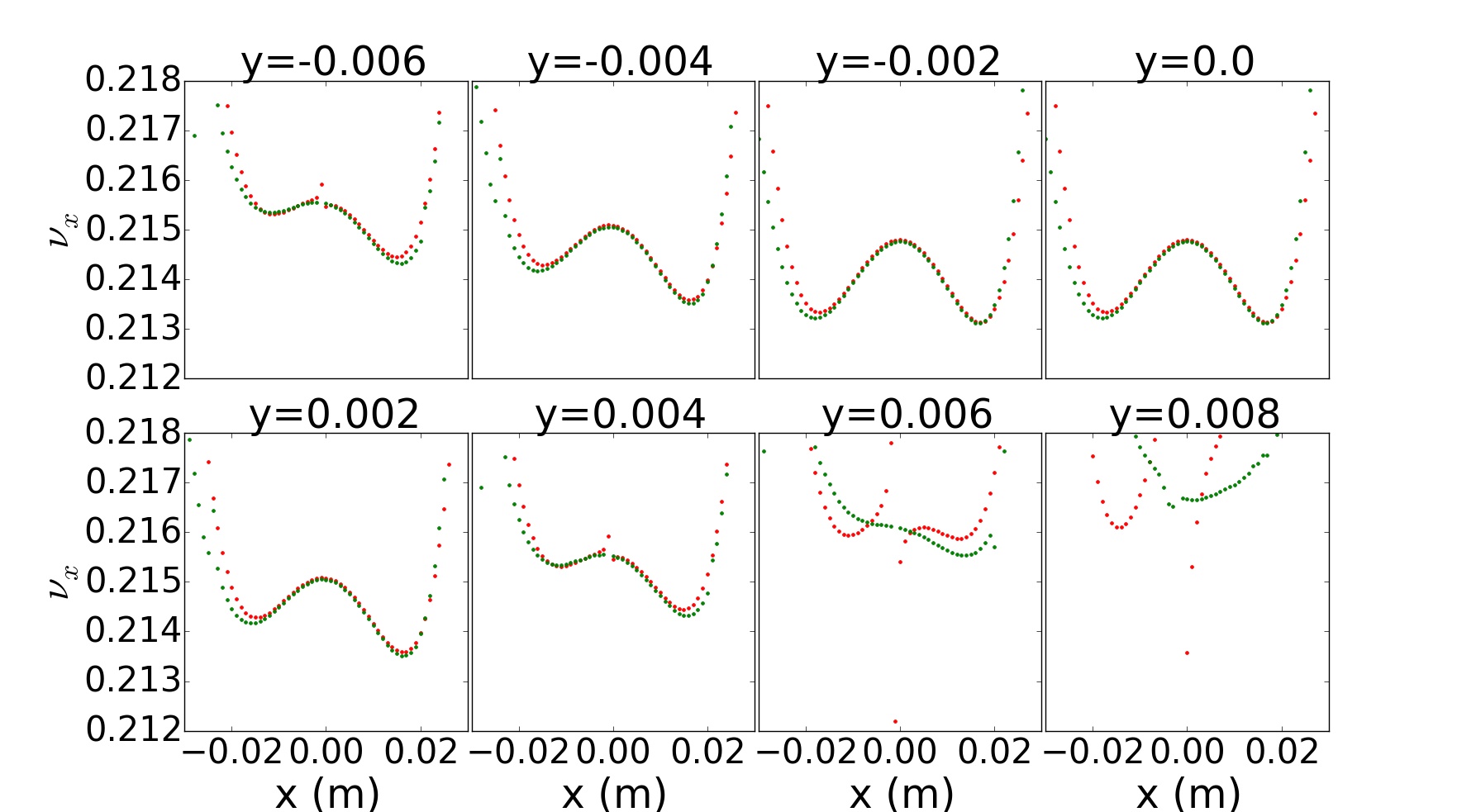}
  \caption{\label{fig:fig9}  Plots for $\nu_x$  as function of x for various y, compare tracking (green) with theory (red)}
\end{figure}

\begin{figure}[htb]
  \centering
\includegraphics[width=\columnwidth]{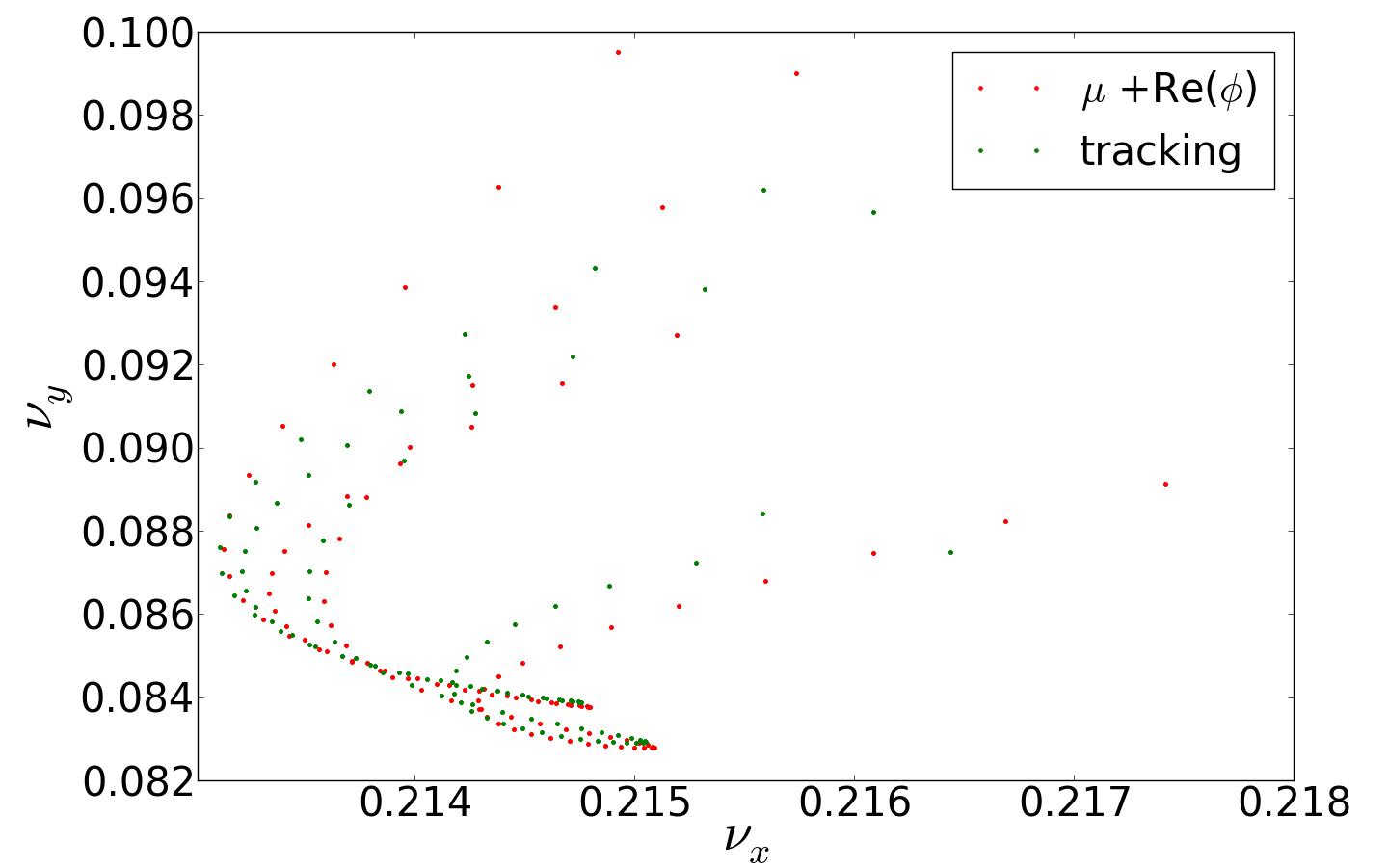}
  \caption{\label{fig:fig10}  Tune footprint from tracking (green) and theory (red)}
\end{figure}

In Fig.10, we plot the tune footprint calculated from tracking (green) and from $\mu+\text{Re}\phi$  . Clearly this shows we can calculate tune footprint approximately from square matrix without the very time consuming tracking particles for various initial x and y.

\subsection{Coherence Region and Dynamic Aperture}
We are interested in the range of the region where our “coherence” conditions Eq.\,\eqref{eq:Eq.1.17} to Eq.\,\eqref{eq:Eq.1.19} are valid. We can find this range by tracking particles with different initial conditions and find the tune variation such as in the calculation for a frequency map\cite{laskar_2003}. However, it is possible to find this range without tracking particles for many turns. For this we need to calculate $\phi$ and $\Delta$  for a set of points where $|w_{x}|$,$|w_{y}|$  are constants.

For a given set of points on  the $\theta_x,\theta_y$ planes, as shown in the two Poincare sections of constant $|w_{x}|$,$|w_{y}|$ of Fig.8, we find their coordinates $z_x,z_y $  using the inverse function of Eq.\,\eqref{eq:Eq.3.3}. From this set of $z_x,z_y $ we use Eq.\,\eqref{eq:Eq.1.11} to calculate $w_0,w_1,w_2 $. Then we use Eq.\,\eqref{eq:Eq.1.18} and Eq.\,\eqref{eq:Eq.1.19} to calculate $\phi$ and $\Delta$. These results are used to calculate the standard deviation of $\text{Re}\phi$, $\text{Im}\phi$ and $\Delta$ .

In Fig.11 we use color scale to represent the RMS value of $\Delta_x$ in xy plane. For every point on this plane, we find the corresponding $|w_{x}|$,$|w_{y}|$ assuming initial $p_x=p_y=0$. Then, for a set of azimuthal angles $\theta_x$,$\theta_y$ of the corresponding $w_x$,$w_y$ we find the inverse function solution for Eq.\,\eqref{eq:Eq.3.3} and use the result $\bar{z}_x$, $\bar{z}_y$ to calculate the standard deviation for $\Delta_x$ for both x and y motion respectively. At x=-1mm, y=5.5-9mm, we see the resonance behavior discussed regarding to Fig.9.
\begin{figure}[htb]
  \centering
\includegraphics[width=\columnwidth]{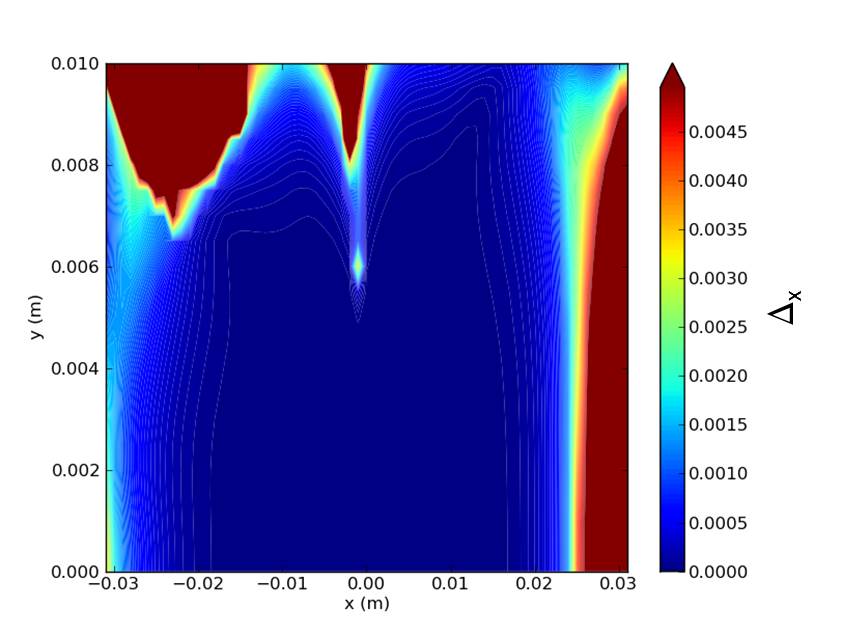}
  \caption{\label{fig:fig11}  $\Delta_x$ in xy plane color scaled by the RMS value }
\end{figure}
For x between 20mm and 25mm and for y from 0 to 8mm, we can see the color changes from dark blue to light blue, passing through yellow to red, reaching dark brown. This is the region where we see $|w_{x}|$ gradually deviates from being a constant during the motion.

In Fig.12, we plot the frequency diagram obtained from tracking using elegant [24] for the same lattice setting.  When compared with Fig.11, we see that it gives a crude picture about the dynamic aperture. Even though, without multi-turn tracking, the plot does not give the detailed structured frequency map, the light blue area gives information about the area of larger tune variation. Fig.11 confirms the expectation that the function $\Delta$ is related to the “coherence” condition or stability condition.

\begin{figure}[htb]
  \centering
\includegraphics[width=\columnwidth]{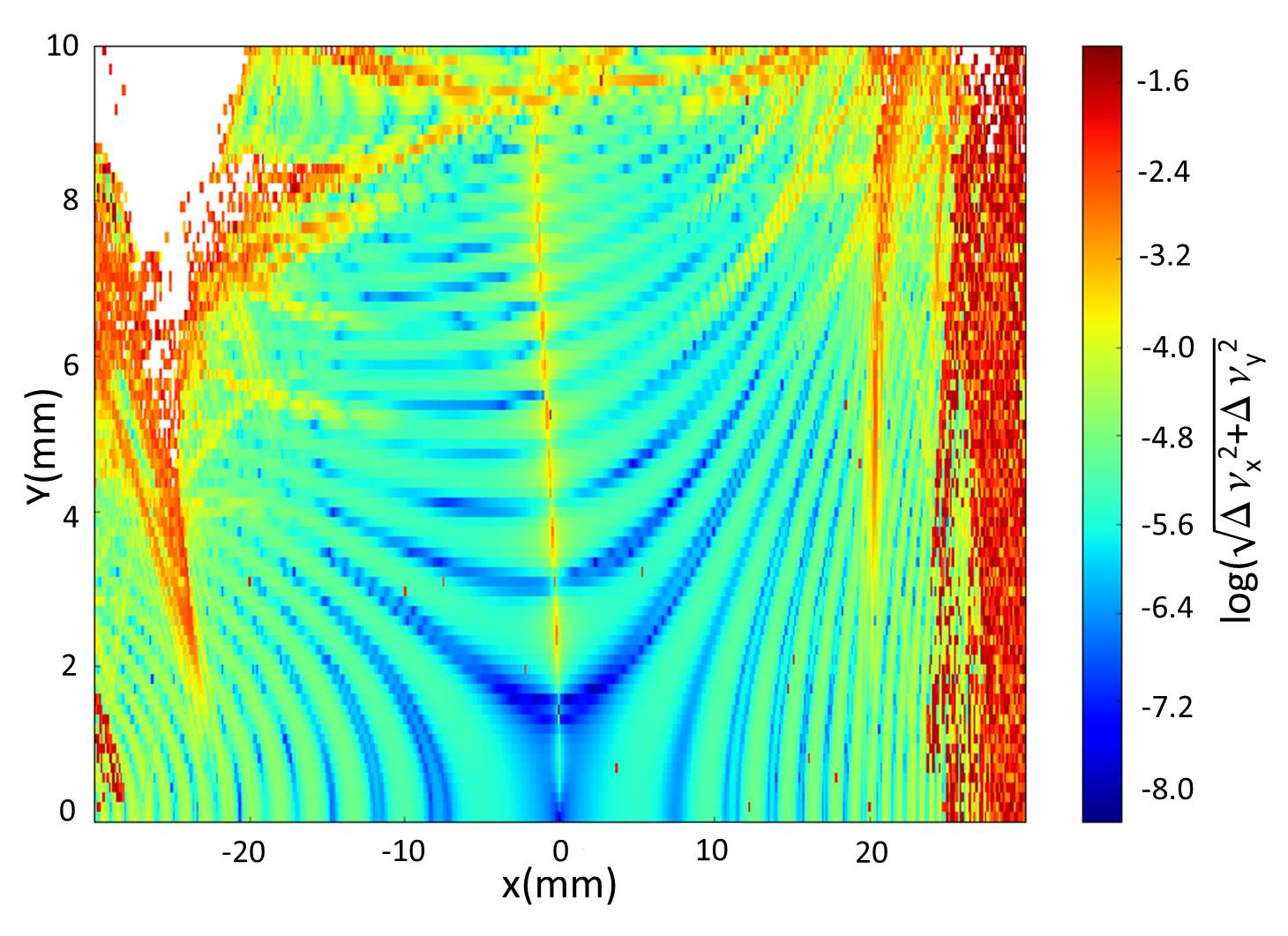}
  \caption{\label{fig:fig12}  Frequency diagram for lattice “nsls2sr\_supercell\_ch77” }
  \setlength{\belowdisplayshortskip}{-4pt}
\end{figure}
\parskip 2pt
\section{One Example of Application: Manipulation of phase space trajectory}
The analysis by the square matrix method given in the previous sections can be used for nonlinear optimization of storage ring lattices. In the following we give an example of phase space trajectory optimization by this method. A separate paper to discuss using the square matrix method to optimize several nonlinear lattices is in preparation \cite{li_2016}.
\begin{figure}[htb]
  \centering
\includegraphics[width=\columnwidth]{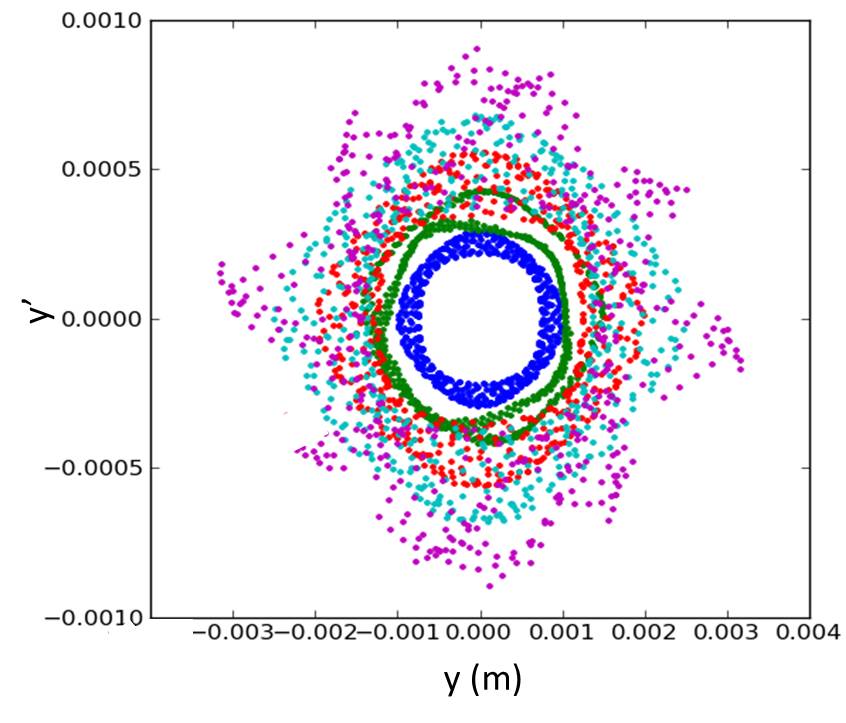}
\includegraphics[width=\columnwidth]{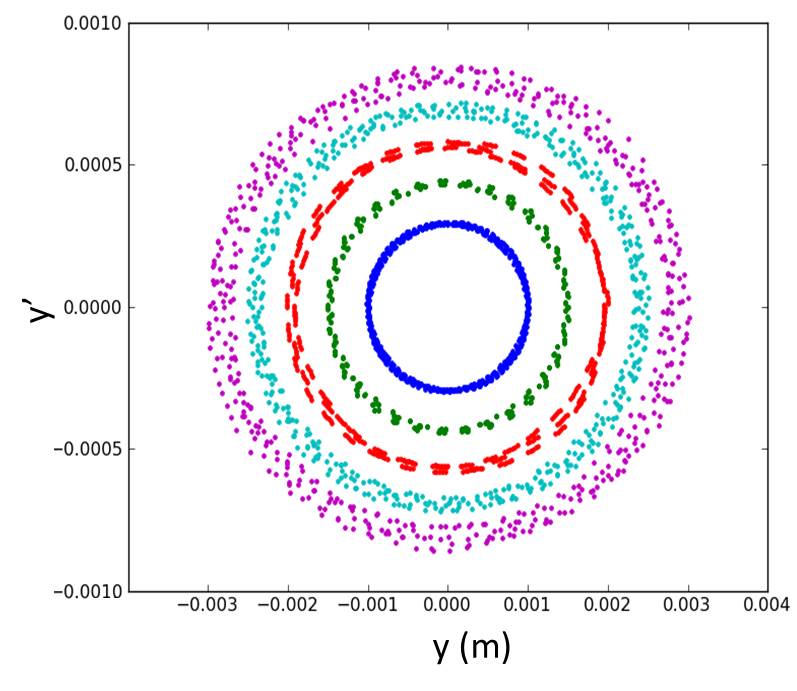}
  \caption{\label{fig:fig13}  Trajectories in y-y’ phase space for 5 particles before (top) and after optimization (bottom) by square matrix}
\end{figure}
As described in Fig.4 and Fig.8, the action defined from Courant-Snyder variable $J_x \equiv |\bar{z}_x |$ and $J_y \equiv |\bar{z}_y |$, as calculated from Eq.\,\eqref{eq:Eq.3.1}, is no longer constant when nonlinearity dominates over linear dynamics. There is a significant distortion from flat planes in the Poincare section. We characterize this distortion by $\Delta J/J=(J_{max}-J_{min})/J_{mean}$. When the distortion is large, the particles receive much larger nonlinear kicks from the higher order multipoles when x reaches maximum, and hence the system becomes more nonlinear. The goal of nonlinear optimization is to reduce the nonlinear distortion, and hence increase the dynamic aperture. In the 1-D case, this is equivalent to optimize the trajectories in the space of Courant-Snyder variables $\bar{x}$,$\bar{p}$ so that they are as close as possible to circles with constant radius.  From our previous analysis in Section 3B, we see that the invariant tori are given by constant $|w_{x}|$, $|w_{y}|$ . Hence, to minimize $\Delta J/J $ , we need to calculate $\Delta J/J $ for contours with constant  $|w_{x}|$, $|w_{y}|$ and vary the sextupoles to minimize $\Delta J/J $ on these contours. In other word, for given pair of $ r_x =| w_x |, r_y=| w_y|$, and for a set of $ \theta_x$,$ \theta_y$ we need to calculate the corresponding set of $\bar{z}_x$, $\bar{z}_y$ , calculate $\Delta J/J $ , then, based on these steps to minimize $\Delta J/J $. It is clear from this we need to use the inverse function solution of Eq.\,\eqref{eq:Eq.3.3}.

However, even though the inverse function calculation is made easy and fast by the use of the inverse matrix mentioned in Appendix D, we would like to carry out this optimization without the inverse function calculation at all. Therefore we remark here that the task of minimizing $\Delta J/J $ is equivalent to optimize the system so that flat planes in the Poincare sections in $w_x$,$w_y$ space (as shown in Fig.8) are mapped to approximate flat planes in the Poincare sections in the $\bar{z}_x$, $\bar{z}_y$ space, and vice versa. Because $w_x$, $w_y$ have been derived as polynomials of $\bar{z}_x$, $\bar{z}_y$ already, the optimization can be carried out by minimization of $|\Delta w /w|$ in the Poincare sections in $w_x$ $w_y$ space instead. This is as shown in Fig.8 but with the rolls of  $w_x$,$w_y$  exchanged with $\bar{z}_x$, $\bar{z}_y$. Thus for a pair of constants $J_x \equiv |\bar{z}_x |$  and  $J_y \equiv |\bar{z}_y |$, and a set of $\psi_x$   $\psi_y$, we calculate $w_x$   $w_y$, then minimize $|\Delta w /w|$.

We applied this optimization for the lattice "nsls2sr\_sepercell\_ch77" which we have discussed in regard of Fig.8. In Figure 13 we compare the trajectories of several particles in phase space $y-y'$ before (top) and after (bottom) the optimization\cite{li_2016}. Different color represents different initial $x, x', y,$ and $y'$. In these 5 pairs of $x$ and $y$, the initial $y$ is chosen to be proportional to the initial $x$. The maximum initial $x$ is 20mm, so the $x$-motion is nonlinearly coupled into y-motion, generating complicated motion in $y-y'$ plane. It is obvious that even though the lattice of the top plot has been optimized for NSLSII operation with nonlinear driving terms minimized already, the further optimization by square matrix method clearly further reduces the nonlinearity of the system significantly.
For this specific example, 3 sets of Poincare sections are selected to minimize $|\Delta w /w|$. The $J_x,J_y$  for these Poincare sections are derived from the following 3 pairs of initial conditions $x_0,y_0$ = \{2.5e-2, 5e-3\};\{1e-2,2e-3\};\{3.5e-2,3e-3\}, respectively. This choice is not unique. Actually, the question about how many Poincare sections should be used, and how many points in each section are taken, is open for future exploration of very fast optimization method.
\section{Square matrix, its structure and its invariant subspace}
All the applications described in Section 3 and 4 are based on the use of the left eigenvectors equation Eq.(1.7), so that we can apply $U$ for the variable transformation from $z$ to the action-angel variable $w_0$ by $W=UZ$ in Eq.(1.11).
In Section 5 and 6, we shall outline the construction of $U$ by Jordan decomposition of the square matrix $M$. For high order, $M$ has very large dimension. Hence we need very efficient way to calculate $U$ from $M$.

The square matrix $M$ has a special property that it is upper-triangular and all its eigenvalues are its diagonal elements, and hence are precisely known. In this section, we first show that because this special property, in a first crucial step, we can reduce the analysis of the very high dimension matrix $M$ into the analysis of its eigenspace of eigenvalue $e^{i \mu}$ with much lower dimension. In Section 6, we show that the final Jordan decomposition is carried out in this eigenspace, and the dimension of the final matrix $U$ is further lower than this eigenspace. Hence the analysis is greatly simplified.

The first step of the analysis of the square matrix is the counting of the number of terms. For two variables such as x and p, for terms of order k, all the terms has a form of $z^{m} z^{\ast(k-m)}$, with $0 \leq m \leq k$, so the number of terms of order k is k+1. So for order n, we need to count the number of of terms from 1 to n+1 (we count k from 0 to n). The sum is (n+1)(n+2)/2 \cite{chao_95}. Thus for 3rd order it is 10, as shown in the example.

For 4 variables such as $x,p_x,y,p_y$, the numbers of terms in order of n is (n+1)(n+2)(n+3)/6. Hence for order 1, 2, 3, ..., n the number of terms are 4,10,20, ...,(n+1)(n+2)(n+3)/6, respectively. For the square matrix dimension, after summing up these numbers of terms, we find it is (n+1)(n+2)(n+3)(n+4)/24. To save space, we shall not give the derivation of the summation here. But it is very easy to test this result. As an example, when truncated to 7’th order, the matrix dimension for 4 variables is 330. This rapid increase of matrix dimension as the order increases seems to indicate a very fast increased complexity of the problem.

However, as we will show in the following, the point of interest is not the dimension of the full matrix but one of its invariant subspaces with much lower dimension. Here "invariance" means that a vector in a subspace of the full space spanned by $Z$, after multiplied by the matrix $M$, remains in the same subspace. In the example represented by Eq.\,\eqref{eq:Eq.1.5}, for 2 variables and at 3rd order, even though the dimension of $M$ is 10, there are only two independent (generalized) eigenvectors with eigenvalue of $e^{i\mu}$, among the 10 eigenvalues in the list $\{1, e^{i\mu},e^{-i\mu}, e^{2i\mu},1 ,e^{-2i\mu},e^{3i\mu},e^{i\mu},e^{-i\mu},e^{-3i\mu} \}$. These two eigenvectors span an invariant subspace of dimension 2 (see Appendix A). Thus the rotation generated by the matrix is represented by a 2$\times$2 matrix in this subspace, representing the nonlinear dynamics.

To count the number of diagonal elements for $M$ with value as $e^{i\mu}$, when we examine how these elements are generated from Eq.(1.2) to Eq.(1.3), we note that all these elements must come from the monomials in $Z$ in the form of $z (zz^\ast)^m$ so that its coefficient is $e^{i\mu} (e^{i\mu}e^{-i\mu})^m=e^{i\mu}$. For a given maximum number $m \geq 0$, the order $n$ must satisfy $n \geq 2m+1$, hence either $n=2m+1$, or $n=2m+2$. For example, if $m=0$, the order n is 1 or 2. If m=1, then n is 3 or 4. Hence for order n, the number of diagonal elements with eigenvalue $e^{i\mu}$ is m+1, i.e, the integer part of (n+1)/2. For order 1,2,3,4,5,6,7 and for the case of two variables, this gives the dimension of the invariant subspace is 1,1,2,2,3,3,4, respectively.

Same way, in the case of 4 variables $x,p_x,y,p_y$, all the elements with value $e^{i\mu_x}$ must come from the monomials in $Z$ in the form of $z_x (z_xz_x^\ast)^{m_x} (z_yz_y^\ast)^{m_y}$. If within order $n$, the maximum number of $m_x+m_y=m$, then we must have $n \geq 2m+1$, hence either $n=2m+1$, or $n=2m+2$. If $m_y=0$ then $m_x$ can take value of $0,1,2, \dots, m$, i.e., there are $m+1$ terms. If $m_y=1$, then $m_x$ can only takes value from $0$ up to $m-1$, i.e., there are $m$ terms. We continue this counting until for $m_y=m$, then $m_x=0$ so there is only one term. Thus we count all the terms including all possible $m_y$, by summing up from 1 up to m+1. The sum is $(m+1)(m+2)/2$. As an example, let m=3, then n is either $2m+1=7$, or $2m+2=8$. Hence for order 7 or 8, the total number of diagonal terms with value $e^{i \mu_x}$ is $(m+1)(m+2)/2=10$. Thus the eigenspace has dimension 10, as compared with the full dimension 330 of the square matrix $M$.

Notice that among these 10 elements, 4 of them have $m_y=0$, 3 of them have $m_y=1$, 2 of them have $m_y=2$, 1 of them has $m_y=3$. Later in the next section, we find that the Jordan decomposition of this eigenspace separates it into 4 invariant subspace with dimension 4,3,2,1 respectively, adding up to 10. The fact that they have the same structure seems not to be a coincidence, even though we do not have a general proof so far.

More importantly, a great simplification comes from the fact that the matrix is upper-triangular with all its eigenvalues given by its diagonal elements precisely determined by the tune $\mu$ as long as we use the variables $z,z^{\ast}$ instead of $x,p$. As is well known and explained in the appendix A, for triangular matrix, the generalized eigenvectors can be calculated in a simple straight forward way.

\section{Invariant subspaces and Jordan decomposition}
\label{sec:section6}
For a n$\times$n matrix $M$, using Jordan decomposition \cite{Golub_2013}(Golub, p.354), we can find a n$\times$n non-singular matrix $U$ such that
\begin{equation}
   \label{eq:Eq.6.1}
\begin{split}
  &UMU^{-1}=
  \begin{bmatrix}
  U_1\\U_2\\...\\U_k
   \end{bmatrix}M
  \begin{bmatrix}
  \overline{U}_1&\overline{U}_2&...&\overline{U}_k
   \end{bmatrix}\\
&=
\begin{bmatrix}
 N_1  & 0   & ... & 0 \\
 0    & N_2 & ... & 0 \\
 0    & 0   & ... & 0 \\
 0    & 0   & ... & N_k
 \end{bmatrix}\equiv N
 \end{split}
\end{equation}
where the $m_j \times m_j$ matrix $N_j={\lambda}_j I_j +{\tau}_j$  with j={1,2,...,k} is the Jordan block with eigenvalue $\lambda_j$, corresponding to the invariant subspace j of dimension $m_j$ in the n dimensional space of vector $Z$. $I_j$  is the identity matrix of dimension $m_j$, while $\tau_j$  is a superdiagonal matrix of dimension $m_j$:
\begin{equation}
   \label{eq:Eq.6.2}
   \tau_j=
\begin{bmatrix}
 0    & 1   & 0 &... & 0 \\
 0    & 0   & 1 &... & 0 \\
 0    & 0   & 0 &... & 1 \\
 0    & 0   & 0 &... & 0
 \end{bmatrix}
\end{equation}
$U_j, \overline{U}_j$  are $m_j \times n$ and $n  \times m_j $ submatrixes respectively. Eq.\,\eqref{eq:Eq.6.1} leads to the following equations: if $ j \neq l , U_j \overline{U}_l =0$ and $U_j M \overline{U}_l=0$ ; if $j=l$, $U_j \overline{U}_l =I_j$; $UM=NU$. Hence the left eigenvector equation:
\begin{equation}
   \label{eq:Eq.6.3}
U_j (M -{\lambda}_j I )={\tau}_j U_j
\end{equation}
(Notice the distinction: $I$ and $I_j$ are identity matrixes for full space $Z$ and the subspace respectively.)

It is seen from these equations that when we study the nonlinear dynamics of the system, we can decompose the motion in the full space $Z$ into the separate motion in many invariant subspaces. In particular, as we showed in Section 3 and 4, the motion in the eigenspace with eigenvalue $e^{i\mu}$    provides a wealth of information about the dynamics. Actually, all other eigenvalues also provide this information. However, for the dynamics of the system of xy motion, we can concentrate on the two simplest invariant subspaces with eigenvalues $e^{i\mu_x}$, $e^{i\mu_y}$ only, hence we drop the index $j$ from Eq.(6.3) from now on.

The Jordan decomposition Eq.(6.1) appears to be complicated because the large dimension of the square matrix $M$. However as pointed out in Section 5, the left eigenspace for one eigenvalue can be separated in one simple step. If we label these left eigenvectors by $e_i$, we find they satisfy the following equation (Appendix A):
\begin{equation}
\setlength{\abovedisplayshortskip}{2mm}
   \label{eq:Eq.6.4}
e_i (M -{\lambda} I )=t_{ik} e_k
\setlength{\belowdisplayshortskip}{2mm}
\end{equation}
They form an invariant subspace, that is, after multiplied by $(M -{\lambda} I )$, any one of them remains to be a linear combination of them represented by the matrix $t$. We used the Einstein convention: the repeated k implies a sum over k. The index i and k run from 1 to m, where m is the null space dimension of the matrix  $( M -{\lambda} I)$. Here $t$ is the matrix derived in Appendix A.

For the example of 4 variables at 7th order, the eigenspace of eigenvalue $\lambda =e^{i \mu_x}$ has dimension 10, and we can find the 10 generalized eigenvectors. $t$ is a $10 \times 10$ upper triangular matrix. Hence our main issue is greatly simplified to finding the Jordan decomposition of $t$. This involves a much smaller amount of work when compared to finding the Jordan form of the matrix  $( M -{\lambda} I )$ itself.

We observe that Eq.(6.4) is very similar in structure to the left eigenvector Eq.(6.3), except that the matrix $t$ is not in Jordan form. As explained in Appendix B, it is easy to carry out Jordan decomposition for a upper-triangular matrix, particularly for the low dimension matrix $t$ with all its diagonal elements zero.

Since all the diagonal elements are zero, once we find the Jordan decomposition $g$, so that $t=g^{-1}\tau g$ with $\tau$ in Jordan form, we find $g_{hi}e_i( M -{\lambda} I )=\tau_{hn}g_{ni}e_i$. Now we can see that the new basis $u_h \equiv g_{hi}e_i$  satisfy the left eigenvector equation $u_h ( M -{\lambda} I )=\tau_{hn}u_n$. Since $\tau$ is in Jordan form with eigenvalue zero only, when we take $u_j$ as the rows of the matrix $U$, this equation is just Eq.\,\eqref{eq:Eq.6.3}.

Hence the eigenspace itself is again separated into several invariant subspaces, all of them have the same eigenvalue zero. About how to find these invariant subspaces, please see Appendix B. For the example mentioned above, the 10 dimensional invariant subspace is separated into 4 subspaces again, with dimension 4, 3, 2, 1 respectively, each is spanned by a chain of generalized eigenvectors. Thus we find the solution Eq.\,\eqref{eq:Eq.6.3} for each of the several subspaces with eigenvalue $\lambda $ without solving for subspace of other eigenvalues.

For simplicity, from now on we drop the index j, and concentrate only on one of the several invariant subspaces of a specific eigenvalue, and we have $U (M -{\lambda} I )={\tau} U $. This equation simply state that every row $u_j$ of $U$ is a generalized left eigenvector of $ (M -{\lambda} I ) $: $u_i ( M -{\lambda}I )=u_{i+1}$ for $0 \leq i<m-1$, and $u_{m-1}$ is the proper eigenvector: $u_{m-1}( M -{\lambda}I )=0$, where m is the dimension of the invariant subspace. All the $u_j$s forms a chain in one invariant subspace. In the previous example, there are 4 chains in the invariant subspace of dimension 10 for the eigenvalue $e^{i\mu_x}$. The lengths of the 4 chains are 4, 3, 2, 1, respectively.

The structure of chains in the invariant subspace of one eigenvalue, with each chain corresponds to one Jordan block, serves as the basis of one of the method of Jordan decomposition. There are many programs available for Jordan decomposition, including some of them providing analytical solution. But occasionally the result is unstable. To ensure stable result, we adopt the method by Axel Ruhe (1970), K\"{a}str\"{o}m and Ruhe (1980a, 1980b)\cite{ruhe_1970, kastrom_1980a, kastrom_1980b}, which is referred to by the text book "Matrix Computations, 4’th Edition" in page 402 of Golub \cite{Golub_2013}. For the convenience of the readers, in Appendix B, we outline the crucial steps of the method without the detailed derivation and the proof of the stability of the decomposition, which is given in these papers.

In order to study the dynamics of the system, as we shall show, the most important information is obtained from the Jordan decomposition of $\text{ln}M$  rather than $M$ itself. As explained in the Appendix C, we can take a logarithm of Eq.\,\eqref{eq:Eq.6.1}, then, we can carry out a second Jordan decomposition of $\text{ln}N$ up to the same order easily and arrive at an equation similar to Eq.(6.1):
\begin{equation}
   \label{eq:Eq.6.5}
\begin{split}
  &U\text{ln}MU^{-1}=
  \begin{bmatrix}
  U_1\\U_2\\...\\U_k
   \end{bmatrix}\text{ln}M
  \begin{bmatrix}
  \overline{U}_1&\overline{U}_2&...&\overline{U}_k
   \end{bmatrix}
  = N
 \end{split}
\end{equation}
Here we redefined the transformation matrix and Jordan form as $U$ and $N$ again to avoid cluttering of notations. After we replace Eq.\,\eqref{eq:Eq.6.1} by Eq.\,\eqref{eq:Eq.6.5}, the formulas following Eq.\,\eqref{eq:Eq.6.1} remain the same except the eigenvalue $e^{i\mu}$ is replaced by $i\mu$  so that now $N_j=i \mu_j I_j +{\tau}_j$ if we are interested in the Jordan block j with tune $\mu_j=m_x \mu_x +m_y \mu_y$ .

Since we are mostly only interested in the analysis of the Jordan blocks with tune either $\mu_x$ or $\mu_y$,  we drop the index j in $U_j$ and similar to Eq.\,\eqref{eq:Eq.6.3} we get
\begin{equation}
   \label{eq:Eq.6.6}
   \begin{split}
&U \text{ln}M =(i \mu I +{\tau})U\\
 \end{split}
 \end{equation}

Notice now we use $U$ and $\overline{U}$ to represent the submatrix of the transformation matrix. Use $U \overline{U} =I$, we get
\begin{equation}
   \label{eq:Eq.6.7}
   \begin{split}
&U \text{ln}M \overline{U} =i \mu I +{\tau} , \hspace{1cm} \text{and}\\
&UM \overline{U}=e^{i \mu I +{\tau}}.
 \end{split}
 \end{equation}

We write this in the form of left eigenvector equation, following the steps from Eq.(6.1) to Eq.(6.3), we get  the main equation Eq.(1.7)
\begin{equation}
   \label{eq:Eq.6.8}
   \begin{split}
&UM =e^{i \mu I +{\tau}}U  \hspace{2mm}
   \end{split}
\end{equation}
For the example of 4 variables $x,p_x,y,p_y$, at 7’th order, the subspace of eigenvalue $e^{i\mu_x}$  is 4 for the longest chain, the matrix $U$ is a $4 \times 330$ matrix, as we described in Section 1B. For the other 3 shorter chains, the dimension is 3, 2, 1 respectively. We will concentrate on studying the longest chain because it has most detailed information about the nonlinear dynamics. About why the longest chain is important, it will become clear at the end of Section 8 and Appendix E after we explain the structure of the chains.

\section{Multi-turns, Tune and Amplitude Fluctuation}
\label{sec:section7}
As described in Section 1B-1C, with the definition $W=UZ$ and its initial $W_0=UZ_0$,
within the region where the invariant tori remain stable, $W_0$ must satisfy the "coherence" condition, and Eq.(6.8)  leads to
 \begin{equation}
   \label{eq:Eq.7.1}
W= e^{i\mu I +{\tau}} W_0 \cong e^{i (\mu+\phi)} W_0.
  \end{equation}
with tune shift $\phi =-i w_1/w_0$. That is, after each turn, every row $w_j=u_jZ$ in $W_0$ rotates by a factor $ e^{i(\mu+\phi)}$  in their separate complex planes like in a perturbed twist map (see, e.g., \cite{ lieb_1983,Turchetti_1994}), or, behaves like an action-angle variable. For example, let $w_0=r e^{i\theta}$ , then $r=|w_0|$ remains the same like an action variable, while $\theta \rightarrow \theta+\mu+\phi$ after each turn like the angle variable.

Near the border of the survival invariant tori, for example, if the system is near its dynamic aperture or near a major resonance, the condition of Eq.\,\eqref{eq:Eq.1.17} and the condition that $\phi$ is real Eq.\,\eqref{eq:Eq.1.18} are violated slightly, and the Eq.\,\eqref{eq:Eq.1.14} also has increased errors. For convenience we call this situation as a deviation from a ”coherent state”. Hence these conditions provide information about dynamic aperture and resonances.

We now consider the map after n turns. Applying Eq.\,\eqref{eq:Eq.7.1} n times, we obtain
\begin{equation}
\label{eq:Eq.7.2}
W(n)= e^{in\mu I +n{\tau}} W_0 =  e^{in\mu } e^{n{\tau}} W_0,
  \end{equation}
with $W(n=0) \equiv W_0$, and we have moved the constant $ e^{in\mu }$ to the front, dropped the identity matrix $I$ to remind us that it is a constant. Before expanding Eq.\,\eqref{eq:Eq.7.2}, we follow the Dirac notation, let
\begin{equation}
   \label{eq:Eq.7.3}
  |0> \equiv
  \begin{bmatrix}
  1\\0\\0\\...\\0
   \end{bmatrix},
  |1> \equiv
  \begin{bmatrix}
  0\\1\\0\\...\\0
   \end{bmatrix},
|2> \equiv
  \begin{bmatrix}
  0\\0\\1\\...\\0
   \end{bmatrix}, .. ,
|m-1> \equiv
  \begin{bmatrix}
  0\\0\\...\\1
   \end{bmatrix},
\end{equation}
Using the expression Eq.\,\eqref{eq:Eq.6.2} for $\tau$, we find chain $\tau|0>=0,\tau|1>=|0>,\tau|2>=|1>,\tau|3>=|2>,\dots$. And, $\tau^2|0>=0,\tau^2|1>=0,\tau^2|2>=|0>,\tau^2|3>=|1>,\dots$. Hence we have
\begin{equation}
   \label{eq:Eq.7.4}
\setlength{\abovedisplayshortskip}{-1pt}
 \begin{split}
W_0&=w_0|0>+w_1|1>+w_2|2>+\dots+w_{m-1}|m-1> \\
e^{n\tau}&=1+n\tau+\frac{n^2}{2}\tau^2+\dots+\frac{m-1}{(m-1)!}\tau^{m-1}
   \end{split}
\setlength{\belowdisplayshortskip}{-1pt}
\end{equation}
\begin{equation}
\setlength{\abovedisplayshortskip}{-1pt}
   \label{eq:Eq.7.5}
   \begin{split}
e^{n\tau}W_0&= \\
w_0|0>&+w_1|1>+w_2|2>+..+w_{m-1}|m-1> +\\
n (w_1|0&>+w_2|1>+..+w_{m-1}|m-2> +...)+\\
\frac{n^2}{2} (w_2|&0>+w_3|1>+..+w_{m-1}|m-3>+...)+..
   \end{split}
\end{equation}
Thus we find
\begin{equation}
   \label{eq:Eq.7.6}
   \begin{split}
W(n)&=  e^{in\mu } e^{n{\tau}} W_0=\\
e^{in\mu }&(w_0+ n w_1+\frac{n^2}{2} w_2+\dots)|0>+\\
e^{in\mu }&(w_1+ n w_2+\frac{n^2}{2} w_3+\dots)|1>+\dots
   \end{split}
\end{equation}

Compare this with the definition of $W(n)$ , we find
\begin{equation}
   \label{eq:Eq.7.7}
   \begin{split}
w_0(n)&=  e^{in\mu }(w_0+ n w_1+\frac{n^2}{2} w_2+\dots)=\\
&e^{in\mu }w_0(1+ n \frac{w_1}{w_0}+\frac{n^2}{2} \frac{w_2}{w_0} +\dots)=\\
&e^{ in\mu+\text{ln} (1+ n \frac{w_1}{w_0}+\frac{n^2}{2} \frac{w_2}{w_0} +\dots)} w_0=\\
&e^{ in\mu+ n \frac{w_1}{w_0}+\frac{n^2}{2} (\frac{w_2}{w_0}-(\frac{w_1}{w_0})^2)+\dots} w_0 \equiv \\
&e^{ in\mu+ i n \phi+\frac{n^2}{2} \Delta +\dots}w_0.
  \end{split}
\end{equation}
Here we used $\phi$ and $\Delta$ given in Eq.(1.17) and Eq.(1.19),
\begin{equation}
\label{eq:Eq.7.8}
i\phi \equiv \frac{w_1}{w_0} ; \Delta \equiv \frac{w_2}{w_0}-(\frac{w_1}{w_0})^2.
  \end{equation}
To avoid cluttering of symbols, all $w_j$  without specification of n here represent $w_j(n=0)$. When compare with Eq.\,\eqref{eq:Eq.7.1}, we identify $\text{Re}\phi$  as the amplitude dependent tune shift. In the region in the phase space where the invariant tori survive with stable frequency, we recognize that $\phi$ is real and remains to be a constant along a circle with radius $r=|w_0|$.  In addition, the term with $\Delta$ in the exponent of Eq.(7.7) is proportional to $n^2$ instead of $n$. Hence it represents the fluctuation of frequency from turn to turn. $\Delta$ should be nearly zero in order for the system to have a stable frequency. For convenience, this was referred to as “coherence condition”:

\begin{equation}
\setlength{\abovedisplayshortskip}{-1pt}
\label{eq:Eq.7.9}
\text{Im}\phi \cong 0 ; \Delta \cong 0 .
\setlength{\belowdisplayshortskip}{4pt}
\end{equation}

We remark that we pay attention only to the first few terms in the exponent of the right hand side of Eq.(7.7) because the terms of higher power of n have factors of $w_m$ with increased m, while as m increases $w_m$ in Eq.(7.7) becomes small and lost information.

The analysis given by Eq.\,\eqref{eq:Eq.7.7}- Eq.\,\eqref{eq:Eq.7.8} paints a physical picture about why and how a chain represents a rotation in the phase space: we need not only $w_0$ to represent a rotation, we also need  $w_1$ and  $w_2$ to provide information about the phase advance and how stable the frequency is. The phase space is divided into many invariant subspaces, each represents a rotation through the phase shift generated by a chain. For each eigenvalue, the longest chain provides the most detailed information about the rotation while the shorter chains and their sub-chains represent approximation with only high power terms and less information. The invariant subspaces of different eigenvalues represent the rotation of different harmonics of the system.

Clearly $w_0,\phi$, and $\Delta$ are all functions of initial value of $z,z^\ast$. Or, if we use inverse function of $w_0(z)$ to represent $z$ as function of $w_0$, then $\phi$ and $\Delta$ both are functions of $w_0$. Thus, given initial value of $w_0$ , we can examine whether $\phi$  is constant along a circle with radius $r=|w_0|$, whether it is a real function, and whether $\Delta$  is nearly zero, and obtain the information about whether the initial state is close to the border of the survival invariant tori or near resonance. The deviation of the real part of $\phi$ from a constant is the tune fluctuation, while the imaginary part of $\phi$  gives “amplitude fluctuation”, i.e., the variation of $r=|w_0|$  after many turns. The non-zero $\Delta$  indicates a deviation from “coherent state”, seems to be related to the Liapunov exponents \cite{lieb_1983}( p.298). All of these has been checked by a comparison with simulation, as we discussed in detail in Section 3 and 4.

However, all of these are based on the stability, precision, and uniqueness of the square matrix Jordan decomposition. As this is an issue often raised whenever one starts to talk about Jordan decomposition, we address it in the next section.

\section{Stability, Precision, and Uniqueness of the Square Matrix Jordan Decomposition }
\label{sec:section8}
There are still two issues remain to be addressed about the Jordan decomposition. The first is its stability and precision. The second is about its uniqueness. Both issues involve some details of Jordan decomposition. Hence in this section we only briefly explain how these two issues are resolved, the details are given in Appendix E.

First, about the stability and precision issue of Jordan decomposition, when we want to achieve high precision near the dynamic aperture, we need to use high order square matrix $M$. During the construction and Jordan decomposition of $M$, the coefficients of high power terms may become many orders of magnitude larger or smaller than the low order terms. When the ratio of these high power terms and the linear terms becomes so large that the small terms in the coefficients are approaching the machine precision, we lost information and cannot distinguish small terms near machine zero from true zero, then the Jordan decomposition breaks down.

This problem is resolved by the scaling $z=\bar{z}/s$ mentioned in Section 3A, where $\bar{z}$ is the Courant-Snyder variable. This scaling is used to reduce the ratio between the maximum coefficients of high power terms and linear terms. In the example given in the Appendix E, for a typical 7 order square matrix with 4 variables, the range of the coefficients is reduced from 18 orders of magnitude to between 0.03 and 35. As result the Jordan decomposition achieves stability and high precision.

Second, the Jordan decomposition we discussed so far is not uniquely determined. One obvious example of this is that when we add $u_1$ to $u_0$ in Eq.(2.6), the left eigenvector equation remains correct. We can check that $u_0(M-i \omega_0 I)=u_1; u_1(M-i \omega_0 I)=0$, so they satisfy the left eigenvector equation $U(M-i \omega_0 I)=\tau U$. We immediately see that when $u_0$ is replaced by $u_0+a u_1$ with any arbitrary number $a$, the left eigenvector equation still holds. As we can see in Section 2, this corresponds to add to $w_0$ a term proportional to $w_1$.

The polynomial $w_1$ represents a circle in the complex plane of $z$, it does not carry any information about the nonlinear distortion, while $w_0$ has all the terms of order 1 to 3 and carries much information about the distortion. If $|a|$ is very large, the nonlinear distortion we see in $w_0$ will be dominated by $aw_1$, which carries no information about the nonlinear distortion. This blurs the nonlinear distortion details given by other terms in $w_0$. Hence we need to choose the coefficient $a$ so that this invariant term is zero in $u_0$. As shown in Eq.(2.6), this term, i.e., the column 7 of the first row of $U$ is indeed already 0.

In Appendix E, we study how to remove from $w_{0x}$ the high power terms of form $z_x (z_xz_x^\ast)^{m_x} (z_yz_y^\ast)^{m_y}$, i.e., the terms of form of $z_x$ times an invariant monomial. For the case of 4 variables at 7 order, we find in the longest chain, $w_{0x}=u_{0x}Z$ is a polynomial with power from 1 to 7, $w_{1x}=u_{1x}Z$ has power from 3 to 7, ..., and $w_{3x}=u_{3x}Z$ has only terms of power 7.

Table II of the Appendix E shows the structure of of different chains by showing the terms of the lowest power. From the table, we see that the only polynomial with power from 1 to 7 is $w_{0x}$ in the longest chain. All the other $w_{jx}$ do not have linear term. We also show why those high power invariant terms destroy the information about nonlinear distortion.
The contribution from these terms can always be systematically removed. In doing so, the Jordan decomposition becomes uniquely defined, and the blur caused by these terms is minimized.

\section{Conclusions and Future Work}

We showed that for a nonlinear dynamic system, we can construct a square matrix. Using linear algebra the Jordan decomposition of the square matrix leads to a variable transformation from $z$ to $w_0$. The one turn map expressed in terms of the new variable $w_0$ is exactly equivalent to the original map expressed in terms of $z$ whenever the inverse function $z(w_0)$ exists. However, the map expressed in terms of $w_0$ has only small deviation from a twist map, i.e, it is a perturbed twist map. The deviation is characterized by the fact that on a circle of constant $|w_0|$ in the $w_0$ plane, $\phi$ is not a constant, and $\Delta \neq 0; \text{Im}(\phi) \neq 0$. Hence these quantities provide a tool to study the frequency fluctuation, the stability of the trajectories, and dynamic aperture. Thus the analysis of nonlinear dynamic system can be greatly simplified using linear algebra.

The main feature of the new method is we can achieve high order in one step. This is a significant advantage when compared with canonical perturbation theory and normal form where the calculation is carried out order to order by complicated iteration. We also showed that the stability and precision of the Jordan decomposition is ensured by scaling the variables, and by removing the high power invariant monomial terms.

In Section 3 and 4, we demonstrated that the action-angle variable remains nearly constant up to near the boundary of the dynamic aperture and resonance lines. They successfully reproduce both the correct phase space structure and tune shift with amplitude. In addition, we tested the  several measures of the stability of the trajectories and their tunes such as the criterions $\text{Im}\phi \cong 0; \Delta  \cong 0$. For sufficient high order, we compared the region where these criterions satisfied with the dynamic aperture for realistic lattices in Section 3, showing these measures can be used to find the dynamic aperture.

In summary, the presented theory shows a good potential in theoretical understanding of complicated dynamical system to guide the optimization of dynamical apertures. Using analysis of one turn map to narrow down the searching range of parameter space before the final confirmation by tracking, the new method can significantly speed up the optimization.

The examples given here are limited to 2 to 4 variables. However, the method developed here is general. Hence the application to 6 variables or more should be possible. The inclusion of bunch length and energy spread into consideration of this method may be interesting for high energy accelerator physics. The analysis given here is general, and hopefully may be applied to other areas, for example, nonlinear dynamics in physics and astronomy.

\begin{acknowledgments}
The author would like to thank Dr. Yongjun Li for his many comments, suggestions and discussion on this paper, in particular, for his contribution to the optimization result used in Section 4. We also would like to thank Dr. Lingyun Yang for his support on the use of his program of TPSA to construct the square matrixes. We also would like to thank Dr. Yue Hao for discussion and comments on the manuscript, and for providing TPSA programs to construct the square matrixes. We also thank Dr. G. Stupakov for a discussion on applying the method to nonlinear differential equation.

We also thank Dr. Boaz Nash for the collaboration on the start and early development in the direction of square matrix analysis of nonlinear dynamics.

This research used resources of the National Synchrotron Light Source II, a U.S. Department of Energy (DOE) Office of Science User Facility operated for the DOE Office of Science by Brookhaven National Laboratory under Contract No. DE-SC0012704.
\end{acknowledgments}
\begin{center}
\bf List of Appendixes \rm
\end{center}
\begin{enumerate}[label=\Alph*.]
\itemsep -0.9mm
\item Eigenspace of Triangular Matrix
\item Outline of a Method of Jordan Decomposition
\item Jordan Decomposition of $\text{ln}(M)$
\item Inverse of an Upper-Triangular Matrix
\item Stability, Precision and Uniqueness of the Jordan Decomposition
\item Minimize Higher Power Terms
\end{enumerate}

\appendix
\makeatletter
\renewcommand\section{\@startsection {section}{1}{\z@}%
                                     {-2.5ex \@plus -1ex \@minus -.2ex}%
                                     {1ex \@plus.3ex}%
                                     {\center\normalfont\bfseries }}
\renewcommand{\thesubsection}{\thesection.\arabic{subsection}}
\makeatother
\section{\label{sec: Aeigenspace} Eigenspace of Triangular Matrix}
As pointed out in Section 5, for triangular matrix, the generalized eigenvectors can be calculated in a simple straight forward way. For a specific eigenvalue, the eigenvectors span an invariant subspace. Here we give a brief description of the method to solve for a set of basis for this subspace using an example.

For an eigenvalue $\lambda$, we need to find the non-trivial solutions of the equation $MZ=\lambda Z$. As an example, we study the eigenspace of the matrix $M$ of Eq.\,\eqref{eq:Eq.1.5} for  $\lambda=e^{i\mu}$. Let $m \equiv M - \lambda I$ with $I$ the identity matrix. Given the set of eigenvalues of
$M$ $\{ 1, e^{i\mu}, e^{-i\mu}, e^{2i\mu},1, e^{-2i\mu}, e^{3i\mu}, e^{i\mu}, e^{-i\mu}, e^{-3i\mu} \}$, we see that $m$ is a triangular matrix with its 2nd and 8’th diagonal element equal to zero, the other 8 diagonal elements are non-zeros. Hence there are only two eigenvectors which span the invariant subspace for this eigenvalue.

To save space for this paper and avoid writing large matrix, let us assume that the 2nd and the 5’th diagonal element are zero instead of 2nd and 8’th. Thus to find the first eigenvector, we write
\begin{equation}
   \label{eq:Eq.A1}
  mX_1=
  \begin{bmatrix}
  a&b&*&*&*&*&*\\
  0&0&*&*&*&*&*\\
  0&0&*&*&*&*&*\\
  0&0&0&c&*&*&*\\
  0&0&0&0&0&*&*\\
  0&0&0&0&0&*&*\\
   0&0&0&0&0&0&*\\
   \end{bmatrix}
  \begin{bmatrix}
  x\\1\\0\\0\\0\\0\\0
   \end{bmatrix}=0
\end{equation}
Here we have chosen $X_1$  only has first 2 rows nonzero, thus clearly we only need to choose  to satisfy the first row of the matrix equation (we use $*$ to represent a certain number which there is no need to specify). That is, we have $ax+b=0$, and hence $x=-b/a$. Thus we have solved for the first eigenvector $X_1$.

Next we find the 2nd eigenvector $X_2$. Since the 2nd zero diagonal element is the 5’th, we let only the first 5 rows of $X_2$ to be nonzero, and let
\begin{equation}
   \label{eq:Eq.A2}
  mX_2=
  \begin{bmatrix}
  a&b&*&*&*&*&*\\
  0&0&*&*&*&*&*\\
  0&0&*&*&*&*&*\\
  0&0&0&c&*&*&*\\
  0&0&0&0&0&*&*\\
  0&0&0&0&0&*&*\\
   0&0&0&0&0&0&*\\
   \end{bmatrix}
  \begin{bmatrix}
  x_1\\x_2\\x_3\\x_4\\1\\0\\0
   \end{bmatrix}=t_{21} X_1=t_{21}
   \begin{bmatrix}
  *\\1\\0\\0\\0\\0\\0
   \end{bmatrix}
\end{equation}
$t_{21}$ is a certain number to be determined. Again we let the 5’th row of $X_2$ to be 1. Clearly we only need to find the first 4 rows of $X_2$ to satisfy the equation. The 4’th row is $cx_4+*=0$ , hence $x_4=-*/c$. This process is repeated to find $x_3$. When we proceed to the 2nd row, the diagonal element becomes zero, hence the situation is different. And we find the equation $*x_3+*x_4+*=t_{21}$ , where $x_2$ is absent and can be set to zero. Since $x_3$  and $x_4$ are already determined, this equation now is used to determine $t_{21}$. This process also explains why in Eq.\,\eqref{eq:Eq.A2} we cannot set $t_{21}=0$ and hence $X_2$ is not a proper eigenvector, but a generalized eigenvector: when it satisfies Eq.\,\eqref{eq:Eq.A2}, we have $mX_2 \neq 0$ , but $m^2 X_2=t_{21}mX_1=0$.

Once $ t_{21}$ is determined, since $x_2$  through $x_4$  are already determined, we can proceed to the first row to solve for $x_1$: $ax_1+*x_2+*x_3+*x_4+*=t_{21}*$ because $a$ is nonzero.

All the rows in $X_1$,$X_2$  are rational functions of elements in the matrix $M$, this is general.

We can generalize this procedure to the case with more than two zero diagonal elements in $m$. Without further details, we summarize the result as follows. As long as the system is sufficiently far away from resonance, for example, the minimum value of $|\lambda-e^{in\mu}|$  for all n within the specified order is several orders of magnitude larger than machine zero, there are well-defined zeros in $m$  matrix. Let the number of zeros to be $n_{\lambda}$  , we can find a set of $n_{\lambda}$  (generalized) eigenvectors $X_j$  such that $mX_j=\Sigma_{0 < k<j}t_{jk}X_k$ with $1 \leq  j \leq n_{\lambda}$. Because $mX_j$  remains to be a linear combination of $X_k$, these $X_j$s serve as basis for the invariant subspace of eigenvalue $\lambda$ of dimension $n_{\lambda}$. Using the Einstein convention (the repeated k implies a sum over k), we have $mX_j=t_{jk}X_k$ , where $t$  is a lower triangular matrix with all diagonal elements equal to zero.

The example here is for the right eigenvectors. To calculate left eigenvectors, we can simply transpose the matrix $M$  and find its right eigenvectors as columns as discussed here, and transpose them back to rows.

The result is the set of left eigenvectors $e_i$ such that
$e_i (M- \lambda I) =t_{ik}e_k $, as given in Eq.\,\eqref{eq:Eq.6.4}.

\section{\label{sec: sectionB}\\ Outline of a Method of
Jordan Decomposition}

As pointed out in Section 6, the structure of chains in the invariant subspace of one eigenvalue, with each chain corresponds to one Jordan block, serves as the basis of the method of Jordan decomposition we outline here. Our main issue has been reduced to the Jordan decomposition of the subspace of matrix $m$ in Appendix A, which is represented by the much lower dimension matrix $t$  with eigenvalue zero.

As pointed out in Section 3, this subspace is spanned by vectors of several chains. Each chain has a proper eigenvector at its end. These proper  eigenvectors form the null space of $m$ . Hence the dimension of the null space of  $m$ is equal to the number of chains. In the example following Eq.\,\eqref{eq:Eq.6.4}, because there are 4 chains with lengths 4,3,2,1 respectively, when multiplied by $m$, the 4 proper eigenvectors become zero, hence the null space $N_1$ for $m$  has dimension 4. The chain of length 1 is removed by $m$, so after multiplied by $m$ only 3 chains left, and the null space $N_2$ for $m^2$  has dimension 4+3=7. Continue this way we find the dimension of the null space $N_k$ for $m^k$ is $m_k=$4, 7, 9, 10 for k=1, 2, 3, 4 respectively. Thus if for every k we can find the basis of null space $N_k$  for $m^k$, we can identify all the eigenvectors as follows. Since $m_4$=10, $m_3$ =9 means there is one vector in $N_4$  independent from the basis of $N_3$, if we can find this vector u which satisfies $m^4u=0$ but $m^3 u \neq 0$, we identify this as the first generalized eigenvector in the longest chain because this is the last remaining vector to become zero when we apply the matrix $m$ to all the basis in the subspace in succession. Thus $u, mu$, $m^2u$, $m^3u$ are the basis of the longest chain of length 4. Once we separate these 4 eigenvectors from the subspace of dimension 10, and find the remaining 6 independent 6 vectors, we can repeat this process to find the vector u such that $m^3u=0$ but $m^2 u \neq 0$ and then find and separate the chain of length 3. Clearly this process can be continued until all (generalized) eigenvectors are separated, thus the Jordan basis is solved. If there are two chains with same length, before the last operation of $m$ to nullify the full subspace, there will be two independent vectors left. We can choose any of them to form a chain, and the another to form another chain. So the process described here is general.

\begin{table*}[htp]
\medskip
  \centering
  \begin{tabular}{ | l | c | c|c|c|c|c| }
    \hline
    Power of  $A$: k &0& 1 & 2 & 3& 4&5\\ \hline
    Rank  & 6 & 4 & 2& 1&0&0\\ \hline
    null space dimension of $A^k$: $m_k$=n-rank=6-rank & 0 & 2 & 4& 5&6&6\\ \hline
    number of chains left after multiplied by $A^{k-1}$:   $n_k=m_{k}-m_{k-1}$ & & 2  & 2 & 1& 1&0\\ \hline
    number of chain terminated by $A^{k}$: $n_{k}-n_{k+1}$ & \hspace{2pc}    & \hspace{0.5pc} 0 \hspace{0.5pc} & \hspace{0.5pc} 1 \hspace{0.5pc} & \hspace{0.5pc} 0 \hspace{0.5pc} & \hspace{0.5pc} 1 \hspace{0.5pc} & \hspace{2pc}  \\ \hline
  \end{tabular}
  \caption{Use the ranks of the powers of a matrix $A$ to calculate null space size and chain structure}
  \label{tab:1}
\end{table*}

Thus the Jordan decomposition in this method requires us to find the null space of the powers of $m$ , and also requires us to separate independent vectors which are in one null space but not in the other. Since the $t$ matrix we would like to decompose into Jordan form, described in Section 3 and in Appendix A, is a triangular matrix with all diagonal element equal to zero and with low dimension, there seems to be a simpler way to carry out the Jordan decomposition than what we shall outline in the following. However, before we can systematically find this simpler way, we just take the method given by K\"{a}str\"{o}m and Ruhe \cite{ruhe_1970, kastrom_1980a, kastrom_1980b}, where the steps we mentioned above are carried out by repeated application of singular value decomposition (SVD). We will give a very brief outline of the steps by an example.
\subsection{\label{sec: subsectionB1} Find the dimension of the null spaces of the powers of
a matrix and its chain structure}
Let us assume a matrix
\begin{equation}
   \label{eq:Eq.A1}
  A=
  \begin{bmatrix}
  0&-1&10&29&-\frac{91}{3}&-96\\[0.3em]
  0&0&0&6&-\frac{10}{3}&-45\\[0.3em]
  0&0&0&1&-3&1\\
  0&0&0&0&1&-2\\0&0&0&0&0&0\\0&0&0&0&0&0\\
   \end{bmatrix}
\end{equation}

The dimension $m_k$ of the null space of $A^k$ can be found by finding the rank of $A^k$ using SVD because the dimension of the matrix, subtracted by the number of zero singular value, is equal to its rank. In numerical calculation, we have to specify a lower limit of singular value which is taken to be zero. If our system is such that the minimum nonzero singular value is many orders of magnitude larger than machine zero, this can be carried without ambiguity. As long as the system is not exactly on resonance, this is easily satisfied.  By SVD we find the Table I, where $n_k=m_{k}-m_{k-1}$ is the number of chains left after multiplying by $A^{k-1}$. From this table, we can see that there are 2 chains of length 2 and 4.

\subsection{\label{sec: sectionB2} Find the null space $N_k$ of $A^k$ }
This is carried out by SVD as follows. By SVD we have $A^{(1)} \equiv A \equiv A_1=U_1\Sigma_1 V_1^H$, where H represents Hermitian conjugate. If we choose SVD such that the singular values are arranged in increasing order, then the first 2 (see Table I:$m_1= n_1=2$) singular values are zeros. Then following \cite{ruhe_1970, kastrom_1980a, kastrom_1980b}, we define $A^{(2)} =V_1^H U_1 \Sigma_1 V_1^H V_1= V_1^H U_1 \Sigma_1 =V_1^H A^{(1)}V_1$, which has all zero as its first $m_1=2$ columns, corresponding to the null space $N_1$.  To find the null space $N_2$, we repeat this procedure for the $(n- m_1)\times(n- m_1)=(6-2)$ $\times (6-2)=4 \times 4$ submatrix $A_2$, which is the lower right corner of $A^{(2)}$, as shown in Fig.14.
\begin{figure}[htb]
  \centering
\includegraphics[width=\columnwidth]{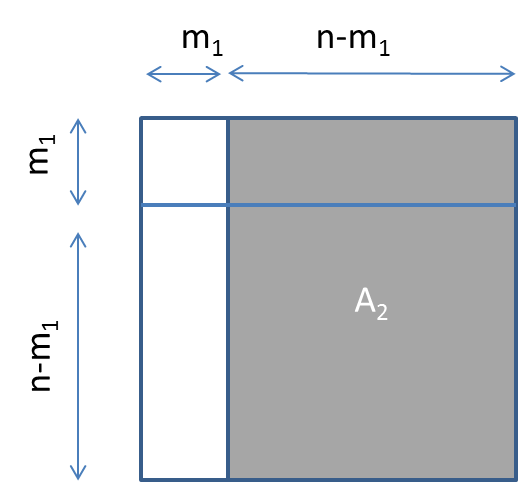}
  \caption{\label{fig:figB1}   Matrix $A^{(2)}$ and its submatrix $A_2$}
\end{figure}

The sequence of unitary transform $V_k$  like this leads to a set of submatrixes $A_k$ each is at the lower right corner of the previous one. The result is a unitary transform $B=W^HAW$ with appearance
\begin{equation}
   \label{eq:Eq.B2}
  B=
  \begin{bmatrix}
  0&B_{12}& B_{13}& B_{14}\\
  0&0& B_{23}& B_{24}\\
 0&0& 0& B_{34}\\
0&0&0& 0\\
   \end{bmatrix}
\end{equation}

Here all the zeros represent blocks of zeros with the left column represent $m_1$ columns of zeros. $B_{ij}$  is a $n_{i} \times n_{j}$ submatrix. According to the table I, the widths of the blocks are 2,2,1,1. Indeed our calculation result agrees with this form.
\begin{equation}
   \label{eq:Eq.B2}
  B=
  \begin{bmatrix}
0&0&-9.99026&-74.19830&15.67463&72.29953\\0&0&0.01922&-23.93292&1.71456&38.34366\\0&0&0&0&-0.40486&6.08191\\0&0&0&0&-0.00032&-0.40472\\0&0&0&0&0&2.19822\\0&0&0&0&0&0
   \end{bmatrix}
\end{equation}
\subsection{\label{sec: sectionB3} Make the Super-Diagonal Blocks of B Upper-Triangular }
The next step is to carry out unitary transform such that each of  $B_{k,k+1}$ blocks is transformed to upper-triangular form. Let
\begin{equation}
   \label{eq:Eq.B2}
  U=
  \begin{bmatrix}
U_1&0&0&0&\\
0&U_2&0&0&\\
0&0&U_3&0&\\
0&0&0&U_4&
   \end{bmatrix},U^H=
  \begin{bmatrix}
U_1^H&0&0&0&\\
0&U_2^H&0&0&\\
0&0&U_3^H&0&\\
0&0&0&U_4^H&
   \end{bmatrix}
\end{equation}
be unitary matrixes with only diagonal blocks nonzero, and the dimension of blocks are $n_{k}$ as in Table I. Let $T=UBU^H$. Then we find
\begin{equation}
   \label{eq:Eq.B2}
  T=UBU^H=
  \begin{bmatrix}
  0&U_1B_{12}U_2^H& U_1B_{13}U_3^H& U_1B_{14}U_4^H\\
  0&0& U_2B_{23}U_3^H& U_2B_{24}U_4^H\\
 0&0& 0& U_3B_{34}U_4^H\\
0&0&0& 0\\
   \end{bmatrix}
\end{equation}
We can choose $U_{4}^H$  to be identity matrix first and carry out QR decomposition $B_{34}=QR$ such that $R$ is upper-triangular and Q is a unitary matrix. Then we choose $U_3=Q^H$, hence  $T_{34}=U_3B_{34}U_4^H=R$ is upper-triangular. We then proceed to $T_{23}=U_2B_{23}U_3^H$. Now since $U_3$ is known, we can carry out another QR decomposition $B_{23}U_3^H=QR$ and choose  $ U_2=Q^H$ so that $T_{23}= R$. Here to avoid cluttering of notation we have repeated the use of the notation $Q$ and $R$ for different matrixes by redefining them each time we use QR decomposition. This procedure continues until $U_1$ is determined to make $T_{12}$ upper-trangular. Then $T$ has the form in Fig.15 (left), with only upper-triangular blocks nonzero and also with all submatrix $T_{k,k+1}$ upper-triangular.
\begin{figure}[htb]
  \centering
\includegraphics[width=4.25cm]{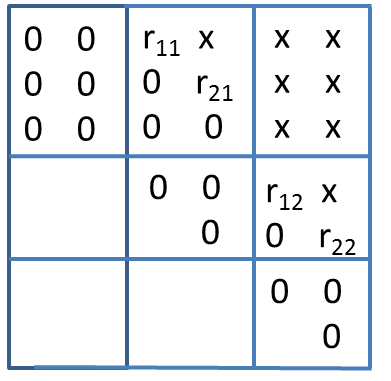}
\includegraphics[width=4.25cm]{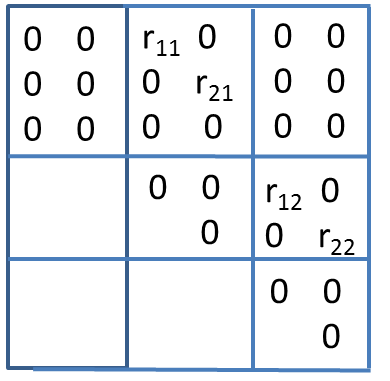}
  \caption{\label{fig:figB2}  The form of T matrix (left) and J (right) }
\end{figure}

For the example, the result agrees with this form:
\begin{equation}
   \label{eq:Eq.B2}
T=
  \begin{bmatrix}
0&0&-10.049&74.190&-15.674&-72.299\\0&0&0&-23.932&1.714&38.343\\0&0&0&0&0.404&-6.081\\0&0&0&0&0&-0.409\\0&0&0&0&0&2.198\\0&0&0&0&0&0
   \end{bmatrix}
\end{equation}

Those elements $r_{ij}$ in the T matrix are called coupling elements. If all elements other than the coupling elements are made zero, then the matrix becomes greatly simplified and the chain structure of the subspace will be revealed, as will become clear in the next subsection.
\subsection{\label{sec: sectionB4} Gauss Elimination by Similar Transformation }
We now proceed to eliminate those elements in $T$ represented by crosses in Fig.15 by Gauss elimination. The procedure is to eliminate all elements other than coupling elements column by column, start from right to left. Consider the similar transform of $T$
\begin{equation}
   \label{eq:Eq.B2}
   \begin{split}
  T&=
  \begin{bmatrix}
0&T_{12}&T_{13}&T_{14}&T_{15}&T_{16}\\
0&0&T_{23}&T_{24}&T_{25}&T_{26}\\
 0&0&0&T_{34}&T_{35}&T_{36}\\
  0&0&0&0&T_{45}&T_{46}\\
  0&0&0&0&0&T_{56}\\
  0&0&0&0&0&0
   \end{bmatrix},\\
   V&=
     \begin{bmatrix}
 1&0&0&0&0&0\\
 0&1&0&0&0&0\\
 0&0&1&0&-a&0\\
 0&0&0&1&0&0\\
 0&0&0&0&1&0\\
 0&0&0&0&0&1
   \end{bmatrix},
 V^{-1}=
     \begin{bmatrix}
 1&0&0&0&0&0\\
 0&1&0&0&0&0\\
 0&0&1&0&a&0\\
 0&0&0&1&0&0\\
 0&0&0&0&1&0\\
 0&0&0&0&0&1
   \end{bmatrix},
   \end{split}
\end{equation}
we find
\begin{equation}
   \label{eq:Eq.B2}
  VTV^{-1}=
  \begin{bmatrix}
0&T_{12}&T_{13}&T_{14}&T_{15}+aT_{13}&T_{16}\\
0&0&T_{23}&T_{24}&T_{25}+aT_{23}&T_{26}\\
 0&0&0&T_{34}&T_{35}&T_{36}-aT_{56}\\
  0&0&0&0&T_{45}&T_{46}\\
  0&0&0&0&0&T_{56}\\
  0&0&0&0&0&0
   \end{bmatrix}
\end{equation}
That is, if we choose $V_{ij}=-a$, then $V_{ij}^{-1}=a$, and the transform simply add the column i multiplied by $a$ to column j, and subtract the row j multiplied by a from row i. Thus if we start from i=4, j=5, and $a=T_{46}/T_{56}$, because the $T$ matrix is upper triangular, $T_{46}$ is eliminated. Column 5 is affected during this transformation, but this does not prevent our elimination process. In particular, the transformed matrix still remains to have the form of the $T$ matrix. When we repeat this procedure with j=5 but let i=3,2,1, the column 6 is eliminated except the coupling element $T_{56}$.

Next we proceed to column 5, but start from i=3, j=4. This eliminates column 5 except the coupling element $T_{45}$. Then, from right to left, the procedure continues to column 4, 3, 2 in the same way. The result is a matrix J in the form shown in Fig.15(right), with only the coupling elements $r_{ij}$ nonzero.

For the example discussed from section B1 to B4, we find
\begin{equation}
   \label{eq:Eq.B2}
  J=UAU^{-1}=
  \begin{bmatrix}
0&0&-10.049&0&0&0\\0&0&0&-23.932&0&0\\0&0&0&0&0.404&0\\0&0&0&0&0&0\\0&0&0&0&0&2.198\\0&0&0&0&0&0
   \end{bmatrix}
\end{equation}
Here we have multiplied all the transformation matrixes used from section B2 through B4 together into one transformation matrix $U$. This matrix is not unitary because in the last steps the matrix $V$ for every step is not unitary, even though in section B2 to B3 all the transformation matrixes are unitary.
\subsection{\label{sec: sectionB5} Permutation and Jordan Form}
The matrix $J$ derived in our example in section B4 indeed only has the 4 coupling elements nonzero: $J_{56}, J_{35}, J_{24}, J_{13}$. It is already very close to the Jordan form. To see the chain structure we let $e_j$ represents the column of 6 elements with only element j equal to 1, and all other elements zero. Then we find
\begin{equation}
\label{eq:Eq.B.10}
\begin{split}
J e_6&=J_{56} e_5, J e_5=J_{35} e_3, J e_3=J_{13} e_1, J e_1=0; \\
 J e_4&=J_{24} e_2, J e_2=0
 \end{split}
  \end{equation}
This confirms the table in B1 that there are two chains of lengths 4 and 2. To transform to standard Jordan form with all the coupling elements equal to 1, we redefine the norm of the basis $e_j$,  and reorder them according to the chain structure, which amounts to a transformation of renormalization followed by a permutation, and we neglect the details here. The result is the Jordan form (we redefine $U$ here to include the renormalization and permutation, and choose to have the longest chain first).
\begin{equation}
   \label{eq:Eq.B2}
  J=UAU^{-1}=
  \begin{bmatrix}
0&1&0&0&0&0\\
0&0&1&0&0&0\\
0&0&0&1&0&0\\
0&0&0&0&0&0\\
0&0&0&0&0&1\\
0&0&0&0&0&0\\
   \end{bmatrix}
\end{equation}
\section{\label{sec: sectionC} Jordan Decomposition of $\text{ln}(M)$}
To find the Jordan decomposition of $\text{ln}(M)$ we first find the transformation for $M$ itself: $UMU^{-1}=N$. Use the property of the similarity transformation, we have $ \text{ln}(UMU^{-1})=U\text{ln}MU^{-1}=\text{ln}N$. Since $N$  is a block diagonal matrix, we have
\begin{equation}
   \label{eq:Eq.C1}
\begin{split}
  &U\text{ln}MU^{-1}=
  \begin{bmatrix}
  U_1\\U_2\\...\\U_k
   \end{bmatrix}\text{ln}M
  \begin{bmatrix}
  \overline{U}_1&\overline{U}_2&...&\overline{U}_k
   \end{bmatrix}\\
  & = \text{ln}N=
\begin{bmatrix}
 \text{ln}N_1  & 0   & ... & 0 \\
 0    & \text{ln}N_2 & ... & 0 \\
 0    & 0   & ... & 0 \\
 0    & 0   & ... & \text{ln}N_k
 \end{bmatrix}
 \end{split}
\end{equation}
For one block with eigenvalue $\lambda_j=e^{i\mu_j}$ we have
$U_j \text{ln}M \overline{U}_j=\text{ln}( e^{i\mu_j}I_j+\tau_j)= \text{ln}(e^{i\mu_j}(I_j+ e^{-i\mu_j}\tau_j))$. Because we are only interested in this invariant subspace, we drop the index j when it is not needed, to write
\begin{equation}
   \label{eq:Eq.C2}
   \begin{split}
  U \text{ln} M \overline{U}=i\mu I+\text{ln}( I+ e^{-i\mu}\tau)= i\mu I+ e^{-i\mu}\tau\\
  -\frac{1}{2} e^{-2i\mu} \tau^2+\dots+\frac{(-1)^{n_j-1}}{ n_j-1} e^{-( n_j-1)i\mu} \tau^{n_j-1}
  \end{split}
 \end{equation}
Here now $U$ represents $U_j$ , i.e., we redefine its submatrix using the same notation to simplify writing. If the dimension of the subspace j is $n_j$, the series terminates at $n_j-1$ because $\tau^{n_j}=0$. The right hand side is not in Jordan form, but it is very simple, and can be transformed by another matrix $V$ (we are excused to redefine the notation $V$ here, not to be confused with the $V$ matrix used in Appendix B4, or Appendix D) into Jordan form
$VU \text{ln}M \overline{U} V^{-1}= i \mu I+\tau $. Now we redefine $VU$ as $U$, $\overline{U} V^{-1}$  as $\overline{U}$, get
\begin{equation}
   \label{eq:Eq.C3}
   U \text{ln}M \overline{U}= i \mu I+\tau
\end{equation}
Notice that since we concentrate only on the subspace of the longest chain in the eigenspace $e^{i\mu_j}$, compared with the discussion in Appendix B, the Jordan decomposition process by $V$ here involves only one chain only.

\section{\label{sec: sectionD} Inverse of an Upper-Triangular Matrix}
The variable transformation of  $w_{0x}(z_x,z_y), w_{0y}(z_x,z_y)$ is defined by the first row $w_{0x}=u_{0x}Z,w_{0y}=u_{0y}Z$ of Eq.(1.11). As explained in Section 3B, very often we need to find the inverse function $z_x(w_x,w_y), z_y(w_x,w_y)$ by solving the equation
\begin{equation}
\label{eq:Eq.D1}
w_x = w_{0x}(z_x,z_y), w_y = w_{0y}(z_x,z_y).
\end{equation}
For this purpose we construct a column $\Phi$ with the rows given by $w_x,w_y,w_x^\ast,w_y^\ast$ and the monomials constructed from them so its transposition $\widetilde{\Phi}$ is similar to the row defined by Eq.\,\eqref{eq:Eq.1.4}.

Now instead of Eq.\,\eqref{eq:Eq.1.3}, we can construct a square matrix $V$
\begin{equation}
\label{eq:Eq.D2}
\Phi=VZ,
  \end{equation}
  where $V$ is also a triangular matrix. It is well-known that the inverse matrix of a triangular matrix is easy to calculate as long as its diagonal elements have no zeros. The linear part of the polynomial is very simple because as $z_x,z_y $  approach zero, they are proportional to $w_x$,$w_y$. We can always choose to multiply $U$ by a constant and divide $\overline{U}$ by the same constant (there are 2 transformation matrixes $U$ for x and y separately but here we are not specific about this point) so that $w_x$,$w_y$ approach $z_x,z_y $  respectively as they approach zero. Thus from Section 1A we find that $V$ has all its diagonal elements equal to 1. Thus it is easy to calculate the inverse matrix $V^{-1}$ which can now be used to calculate $z_x,z_y $ approximately when $w_x$,$w_y$ are given. The result can be used as a set of initial trial values for a more accurate solution of the inverse function of Eq.(D.1). In most our applications the triangular inverse matrix $V^{-1}$ gives an excellent solution already, and there is no need to further improve the precision by solving the inverse function more precisely.

Suppose we want to find the inverse of an upper-triangular matrix $V$ of dimension n. In order to have inverse, all the diagonal elements of $V$ are nonzero. Let us find a matrix $L$  consists of column $y_k$
\begin{equation}
   \label{eq:Eq.D3}
L=(y_1, y_2, y_3,... y_k,..., y_n)
\end{equation}
And we first find  $y_k$ such that
\begin{equation}
   \label{eq:Eq.D4}
   Vy_k=
 \begin{bmatrix}
*&*&*&*&*&*&*&\\
0&*&*&*&*&*&*&\\
0&0&*&*&*&*&*&\\
0&0&0&*&*&*&*&\\
0&0&0&0&a&*&*&\\
0&0&0&0&0&*&*&\\
0&0&0&0&0&0&*&
 \end{bmatrix}
  \begin{bmatrix}
\dots\\\dots\\\dots\\\dots\\x_k\\x_{n-1}\\x_n
 \end{bmatrix}=e_k \equiv
   \begin{bmatrix}
0\\0\\0\\0\\1\\0\\0
 \end{bmatrix}
\end{equation}
where $e_k$  has all rows zero except the k'th row equal to 1. Solving the equation starting from the last row, we find $x_n=x_{n-1}=...=0$ until we reach the k'th row, where we have $ax_k=1$. Hence, $x_k=1/a$. Then we can find $ x_{k-1}$ and continue up to $x_1$. Thus we find $y_k$ has only nonzero rows above and including k'th row. Once we find all the $y_k$, clearly we have the inverse matrix
\begin{equation}
   \label{eq:Eq.D5}
V^{-1}=L=(y_1, y_2, y_3,...y_k,..., y_n)
\end{equation} as an upper-triangular matrix too.

\section{\label{sec: sectionE}Stability, Precision and Uniqueness of the Jordan decomposition}
\subsection{Ensure Stability and Precision by Scaling}
Very often, Jordan decomposition of a matrix is considered to be ill conditioned. However, due to the upper triangular property and because its eigenvalues are precisely known on the unit circle, the matrix $M$ is already in the form of a stable Schur decomposition\cite{Golub_2013}, (Schur decomposition is the step to find the eigenvalues and triagularize the matrix), and its Jordan decomposition is stable except when we are very close to resonance where some eigenvalues are nearly degenerate. When we are sufficiently far from resonance, all the eigenvalues of the matrix $(M-\lambda I)$  other than zeros are sufficiently far from zero so the null space, i.e., the invariant subspace with eigenvalue $\lambda$ can be solved for its eigenvectors to very high precision.

However, we may lose precision during the Jordan decomposition of the matrix $t$  in the invariant subspace as given in Eq.\,\eqref{eq:Eq.6.4} when we use high order square matrixes. Eq.\,\eqref{eq:Eq.6.4} tells us that when $(M-\lambda I)$ acts on the vectors in the invariant subspace of eigenvalue $\lambda$ from right, it is equivalent to the much lower dimension matrix $t$ acts on the vector from the left. Hence in the following discussion we speak of them ($t$ or $(M-\lambda I)$) as if we were talking about the same thing.

The procedure of Jordan decomposition is based on the fact that the invariant subspace is spanned by several chains of eigenvectors, and the end of each chain contributes to the null space of the matrix: $u_{m-1}(M-\lambda I)=0$, as explained in Section 6 and in the second paragraph of Appendix B in particular. Using Eq.\,\eqref{eq:Eq.6.4}, we see that the last right eigenvector (the proper right eigenvector) of $t$ for each chain contributes to the null space of $t$. When the null spaces of different powers of $t$ are calculated, it then becomes easy to build the vectors in each chain and establish the Jordan decomposition, as explained in Appendix B. More details about this procedure are given in Appendix B, and we refer to references \cite{ruhe_1970, kastrom_1980a, kastrom_1980b}.

Here we only point out that our method of Jordan decomposition is based on finding the null spaces of different powers of $t$.  To find the null space correctly we need to distinguish small singular values from zero using singular value decomposition. The high power of $t$ may have very small singular values almost reaching the machine precision. In this case we cannot correctly separate the null space anymore, and the Jordan decomposition fails. Hence we need narrow the range of the singular values so that we can distinguish the minimum singular value from zero clearly.

 We found that the range of singular values depends on the range of the absolute value of the coefficients of monomial terms at different orders in the square matrix $(M-\lambda I)$.  By scaling the variables $z_x=x-ip_x, z_y=y-ip_y$ by a factor $s$, we can reduce the ratio of the maximum and the minimum absolute value of the coefficients above zero. For example, we found for a specific lattice the square matrix has its maximum of the absolute value of the coefficients at 7’th order as $1.38 \times 10^{18}$, and the minimum of those other than zeros is found to be the first order as 1 (the coefficients of order of machine zero are excluded). Then let $s^7 \times 1.38 \times 10^{18}=1 \times s$, we find $s=(1.38 \times 10^{18})^{-\frac{1}{6}}=0.00115$. Now we let $sz_x=x-ip_x, sz_y=y-ip_y$ to make the coefficients of these two terms equal, then we find that the range of the coefficients in the new square matrix using new variables has the coefficients span much smaller range. And the range of singular values is reduced from 18 orders of magnitude to between 0.03 and 35 after the scaling. In Fig.16 we show the spectrum of the singular values before (red) and after the scaling (blue). Therefore, the null space of the invariant subspace is clearly identified; the Jordan decomposition is very stable and accurate.
\begin{figure}[htb]
  \centering
  \includegraphics[width=\columnwidth]{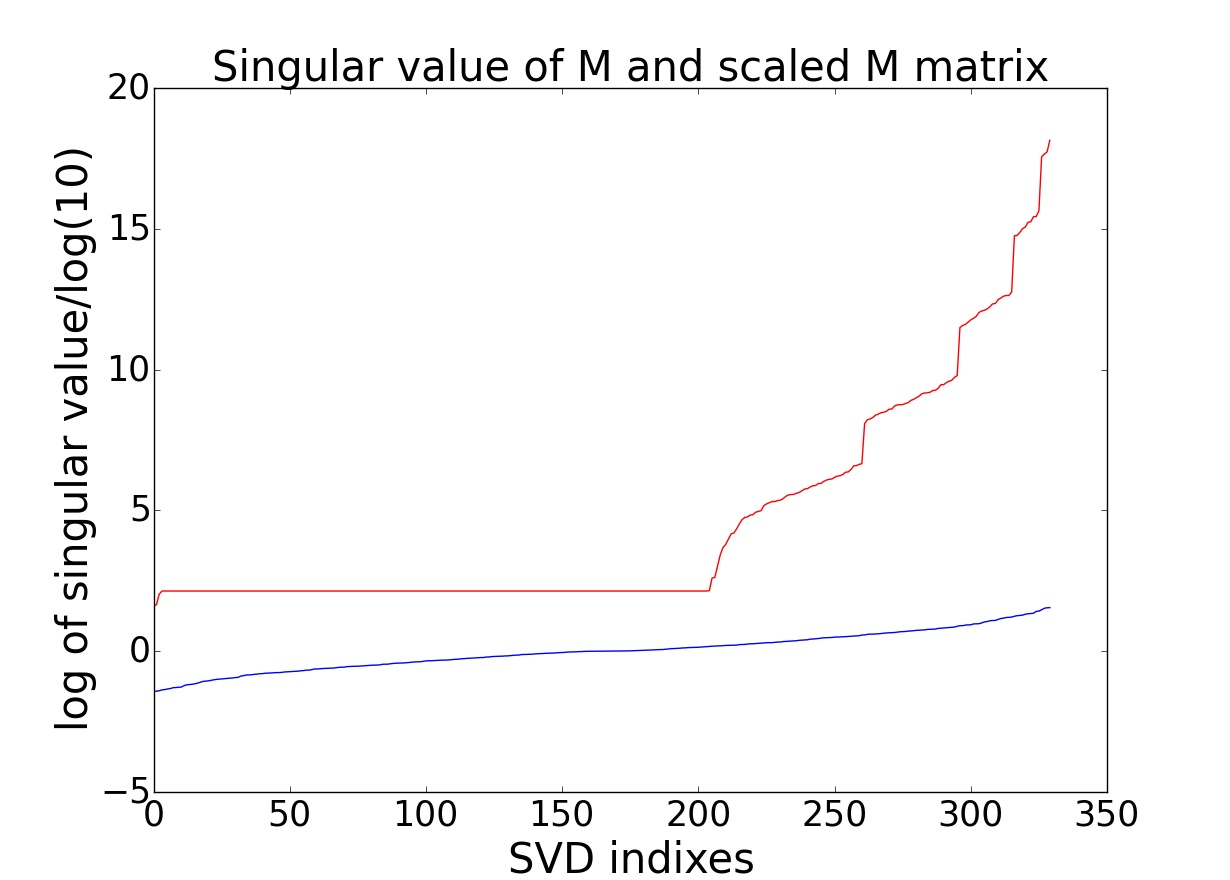}
  \caption{\label{fig:fig16} The range of singular value of the square matrix before (red) and after scaling (blue)}
\end{figure}\\

Another possible loss of precision comes from the construction of the square matrix because the coefficients of the high power terms in the Taylor series expansion in equations similar to Eq.\,\eqref{eq:Eq.1.1} are either larger or smaller by many orders of magnitudes than the linear terms. Thus when we sum over these terms, the small terms lost precision (the effective number of digits) because the limitation on the number of decimal points. This problem can also be solved by scaling method same as mentioned above, the only difference is instead of consider the coefficients of the square matrix, we consider the coefficients in the Taylor expansion in Eq.\,\eqref{eq:Eq.1.1}.  The result not only reduced the error of the square matrix, it also reduces the range of the singular values of the matrix $(M-\lambda I)$. The improvement of the scaling before the construction of the square matrix is so significant that in all the cases we studied, there is very little improvement from the second scaling based on the coefficients of the full square matrix. However, since the second scaling is very simple, and does provide some improvement over the maximum-minimum singular value ratio, we always carry out a second scaling.

When on resonance, the eigenvalue $e^{i  \mu_x}, e^{i \mu_y}$ is degenerate to the eigenvalue of other harmonic of the tune, in the form of $e^{i(m \mu_x+n \mu_y)}$. The structure of the Jordan block is different, and the resonance issue should be treated separately, and is not discussed in this paper. However, as our numerical examples show, in Section 3, our present analysis of nonlinear dynamics by square matrix method is valid even when it is quite close to a resonance.

\subsection{Non-uniqueness of Jordan decomposition}
Another issue of Jordan decomposition is whether it is not unique, and whether we need to impose other conditions based on physics to make it unique. The non-uniqueness of Jordan decomposition is due to its chain structure. A simple example is a chain of two vectors $u_0$ and $u_1$, with $Mu_0=u_1$ and $Mu_1=0$. Then it is obvious that another two vectors $u'_0=u_0+au_1$ and $u'_1=u_1$ also satisfy the same chain relation: with $Mu'_0=u'_1$ and $Mu'_1=0$, where $a$ is an arbitrary constant. Hence the basis of Jordan decomposition is not uniquely defined, and the question now is what kind of conditions based on physics will allow us to determine the choice of $a$. To solve this problem, we must first understand the specific structure of the chains.

\subsection{The structure of the chains in the Jordan decomposition of nonlinear dynamics square matrix}
In the case of two variables x and p, we find only one chain for one eigenvalue $e^{i  \mu}$  . For example, for 7’th order, the invariant subspace we found has 4 (generalized) eigenvectors $ u_0,u_1,u_2,u_3$ as the basis of the space. We find that $w_0=u_0Z$  is a polynomial with powers from 1 to 7, $w_1=u_1Z$  has powers from 3 to 7, $w_2=u_2Z$  has powers from 5 to 7, $w_3=u_3Z$   has only a term of power 7, which is proportional to $z(zz^\ast)^3$. When the matrix $ (\text{ln}M-i\mu I)$ operates from right on the invariant subspace of $Z$ spanned by $u_0,u_1,u_2,u_3$, its operation in the representation using $u_0,u_1,u_2,u_3$ as basis, is given by the simple matrix $\tau$ (see Eq.\,\eqref{eq:Eq.6.6}). Using the equations of left eigenvectors $U(\text{ln}M-i\mu I)=\tau U$, we have $u_0(\text{ln}M-i\mu I)=u_1$, $u_1(\text{ln}M-i\mu I)=u_2$, $\dots, u_3(\text{ln}M-i\mu I)=0$, i.e., they form a chain. Intuitively, the following helps to understand why when multiplied by the matrix $ (\text{ln}M-i\mu I)$, the lowest power terms of the eigenvectors increase progressively by power of 2: the off-diagonal terms in the matrix has at least power of 2.

This is for two variables x and p. For 4 variables such as $x,p_x,y,p_y$, as in the case we studied for a storage ring, there are more independent chains than one for the invariant subspaces of both eigenvalues $e^{i\mu_x}, e^{i\mu_y}$. For example, at 7’th order, for the eigenvalue $e^{i\mu_x}$, in addition to the chain $ u_0,u_1,u_2,u_3$, there is a linear independent chain $u^{\prime}_0, u^{\prime}_1, u^{\prime}_2$, another chain $ u''_0,u''_1$ and a chain with only one element $u'''_0$, forming an invariant subspace of dimension 10. There is only one lowest power term in $u_0$:  $z_x=x-ip_x$. We list the lowest power terms of the chains in this example in Table II.
\begin{table}[htb]
  \caption{ Terms with lowest power in chains}
  \medskip
  \centering
  \begin{tabular}{|c| c| c| c| c|}
  \hline
   Lowest power  &\ 1 \ & 3 & 5  & 7   \\
\hline
   chain number   &  &  & &\\
  \hline
 1  &\ $ u_0$\ &$u_1$&$u_2$&$u_3$\\
  \hline
 2  &      &$ u'_0$&$u'_1$&$u'_2$\\
  \hline
 3  &      &       &$u''_0$&$u''_1$\\
  \hline
4   &      &       &       &$u'''_0$\\
  \hline
 &\ $z_x$\ & $z_x(z_xz_x^\ast)$&  $z_x(z_xz_x^\ast)^2$&$z_x(z_xz_x^\ast)^3$  \\
   Terms of& & $z_x(z_yz_y^\ast)$&  $z_x(z_xz_x^\ast)(z_yz_y^\ast)$&$z_x(z_xz_x^\ast)^2(z_yz_y^\ast)$  \\
Lowest power  &  &                  &  $z_x(z_yz_y^\ast)^2$&$z_x(z_xz_x^\ast)(z_yz_y^\ast)^2$  \\
 &           & &                  &             $z_x(z_yz_y^\ast)^3$  \\  \hline
  \end{tabular}
  \label{table:table2}
\end{table}

Thus the pattern appears: the lowest power terms in the first vector of a chain or sub-chain are always terms with $z_x$ times the powers of the invariant monomial $z_xz_x^\ast$ or $z_yz_y^\ast$  so that for small amplitude, it represents a simple rotation as $z_x$. The first vectors of the shorter chains have terms with $z_x$ multiplied by higher powers of these invariants. This feature helps us to understand its effects on the non-uniqueness of the Jordan decomposition.
\subsection{Physical meaning of non-uniqueness}
To understand the meaning of the non-uniqueness, let us assume $w_0=z+z^3=z(1+z^2), w_1=z(zz^\ast)$. That is the simplest case where the end of the chain is $z$ multiplied by an invariant factor of $zz^\ast$. When $|z| <<1$, for a circular motion in z-plane, both $w_0$ and $w_1$  correspond to a circular motion in w-plane.  But as z increases, $|w_0|$ no longer remains constant because as the phase of $z^2$ changes the factor $1+z^2$ has interference between its two terms. This modulation of amplitude gives distortion of the trajectory, because a constant $|w_0|$ does not correspond to a circular trajectory in z-plane any more. This means $w_0$ carries information about the distortion of the trajectory, while $w_1$ does not carry this information. When $w_1$ is multiplied by a large number and added to $w_0$, it dominates over $w_0$ , and the information about distortion lost.

It is clear now that those terms of $z$ times the powers of the invariant of form $zz^\ast$ represent pure circular motion in phase space and they do not present any information about nonlinearity. Therefore when they are mixed into the first vector of the longest chain, they blur the distortion generated by the interference between the linear term and terms of other harmonics such as $z^3$, and the result is the non-uniqueness.
\subsection{The need to minimize high power terms from the first vector of the longest chain}
In Table II, terms such as $z_x^3$ can appear in $u_0$ but it is not in the lowest power terms in $u_0$ because if it is then it will create a tune $3\mu_x$ rather than $\mu_x$ even for small amplitude. These terms such as $z_x^3$  interferes with the dominating first order term $z_x$ in $u_0$ to generate distortion. A mixture of long chain with short chains in Table II such as $u_0^{(1)} \equiv u_0+au_1+bu'_0, u_1^{(1)} \equiv u_1+au_2+bu'_1, u_2^{(1)} \equiv u_2+au_3+bu'_2, u_3^{(1)} \equiv u_3$, is still a chain. If $a$ or $b$ are very large, the distortion generated by $z_x^3$ will be dominated over by other high power terms and the trajectory becomes more close to a circle with lost information. Hence we need to choose $a$ or $b$ to minimize the high power terms.
It is obvious that a polynomial with many terms should be able to describe much more detailed complicated curve or surface than a single monomial. Also, it is easy to understand that we need to minimize high power terms to increase the convergence radius of a Taylor series.

Therefore, it becomes clear, to extract more detailed information from $u_0$ we need to find a way to minimize higher power terms while maintain a chain satisfying the left eigenvector relation in the invariant subspace. Thus in the Appendix F, we study and find the linear combination of sub-chains such that the higher power terms are minimized. For example, the sub-chain $ u_1,u_2,u_3$ is used to minimize the 3rd power terms in $u_0$ , the sub-chain $ u_2,u_3$ is used to minimize the 5th power terms in $u_0$ , and the sub-chain $u_3$ is used to minimize the 7th power terms in $u_0$.

We remark here it is observed that in the Table II the number of chains for each order happen to be equal to the number of lowest power terms, so that the linear combination of these chains can be used to remove these terms from $u_0$ completely.
Hence when all these higher power terms are minimized, it amounts to removing all the terms of form of $z_x$ times the powers of the invariants, this is equivalent to separate the shorter chains from the longest chain.
The linear combination is uniquely determined, and hence the Jordan decomposition is stable, accurate, and unique. The procedure is described in Appendix F.

The terms we discussed here are among the nonlinear driving terms, the number of which is much more than we list in the table. With limited number of sextupoles, it is impossible to make them all zero, so they can only be minimized in the process of optimization of the dynamic aperture. However, in our description of the chain structure, these terms are all connected in the longest chain. In other words they are correlated. Hence the required number of parameters to be varied is largely reduced.

\section{\label{sec: sectionF} Minimize Higher Power Terms}
As explained in Section 8 and Appendix E, we would like to minimize the higher power terms in the first vector of the longest chain by adding linear combination of sub-chains to the longest chain. This is a change of basis in the invariant subspace of a specific eigenvalue, i.e., a linear coordinate transformation. We use an example to illustrate the transformation.

Let us assume the matrix $N$  in Eq.(6.1) has two chains of length 4 and 3. We study a similar transformation $t$  which changes the basis of the eigenspace but keeps the Jordan matrix $N$ invariant. Thus we write
\begin{equation}
   \label{eq:Eq.F1}
 N=
 \begin{bmatrix}
 A &0\\
 0&B
 \end{bmatrix};\hspace{2pt}
  t=
 \begin{bmatrix}
 I_1+T_1 &T_2\\
 0&I_2
 \end{bmatrix};\hspace{2pt}
  t^{-1}Nt=N
\end{equation}
where $A$ and $B$ are of form Eq.(6.2) with only superdiagonal elements equal to 1 and all other elements equal to zero. $A$ and $I_1$  have dimension 4 , $B$ and $I_2$ have dimension 3 respectively. The matrix $t$ is in a general form to represent a change of basis so that the longest chain after the transformation becomes a linear combination of the long chain from the 4$\times$4 matrix $T_1$  and the short chain from the 4$\times$3 matrix $T_2$. In Appendix E, we mentioned that the Jordan form is not unique. But from Eq.\,\eqref{eq:Eq.F1} we can obtain all the generally possible Jordan basis for the longest chain. Our goal is to choose $T_1$ and $T_2$ to minimize the higher power terms in the first vector in the basis for matrix $N$, as we explained in Appendix E.

Eq.\,\eqref{eq:Eq.F1} leads to $ Nt=tN$, thus we have
\begin{equation}
   \label{eq:Eq.F2}
 \begin{bmatrix}
 A+AT_1 &AT_2\\
 0&B
 \end{bmatrix}=\hspace{2pt}
 \begin{bmatrix}
  A+T_1A &T_2B\\
 0&B
 \end{bmatrix},\hspace{2pt}
\end{equation}
i.e., $AT_1=T_1A$ and $T_2B=AT_2$. To determine the form of the matrix $T$, we examine the effect of Jordan form on column and row. Let
\begin{equation}
   \label{eq:Eq.F3}
u=
 \begin{bmatrix}
 e_1\\ e_2\\e_3\\e_4
\end{bmatrix};v=
 \begin{bmatrix}
  e_1& e_2& e_3& e_4
 \end{bmatrix};A=
\begin{bmatrix}
 0& 1& 0&0\\ 0& 0& 1&0\\0& 0& 0&1\\0& 0& 0&0
 \end{bmatrix};
\end{equation}
we find
\begin{equation}
   \label{eq:Eq.F4}
Au=
 \begin{bmatrix}
  e_2\\e_3\\e_4\\0
\end{bmatrix};vA=
 \begin{bmatrix}
 0& e_1& e_2& e_3
 \end{bmatrix}.
\end{equation}
Hence when acted from left by the Jordan matrix, the column shifts up, leaving the last row zero; when acted from right, a row shifts to the right, leaving the left column zero. Thus $AT_1=T_1A$  means the matrix $T_1$ after shifted up should be the same as it is shifted to the right. Examine this pattern, we see that $T_1$  must be upper triangular, and all the elements on the same superdiagonal must be equal to each other. Also, $T_2$  is of this form. So let
\begin{equation}
   \label{eq:Eq.F5}
t=
 \begin{bmatrix}
1&x_1&x_2&x_3&x_4&x_5&x_6\\0&1&x_1&x_2&0&x_4&x_5\\0&0&1&x_1&0&0&x_4\\0&0&0&1&0&0&0\\0&0&0&0&1&0&0\\0&0&0&0&0&1&0\\0&0&0&0&0&0&1
\end{bmatrix};\hspace{2pt} u=
 \begin{bmatrix}
e_1\\e_2\\e_3\\e_4\\e_5\\e_6\\e_7
 \end{bmatrix}.
\end{equation}
We find $u' \equiv tu=$
\begin{equation}
   \label{eq:Eq.F6}
 \begin{bmatrix}
e_1+e_2 x_1+e_3 x_2+e_4 x_3+e_5 x_4+e_6 x_5+e_7 x_6\\e_2+e_3 x_1+e_4 x_2+e_6 x_4+e_7 x_5\\e_3+e_4 x_1+e_7 x_4\\e_4\\e_5\\e_6\\e_7
 \end{bmatrix}.
\end{equation}

$u'$ is the basis of the matrix $N$ after the transform. Our goal is to minimize the high power terms in the first vector of $u'$, i.e., $e'_1=e_1+e_2 x_1+e_3 x_2+e_4 x_3+e_5 x_4+e_6 x_5+e_7 x_6$ in the basis. If the lowest power term in $e_2$ is 3rd power, as explained in Appendix E, we need to minimize the 3rd power terms of $e'_1$ by varying $x_1$. We first calculate $f=( (e'_1)^\ast e'_1 )_3$. The subscript 3 means we take only the 3rd power terms in the scalar product, hence f here is considered to be the norm of the vector $e'_1$ at 3rd power. The minimization of f requires $\partial f/ \partial x_1=(e'_1)^\ast e_2=0$. Taking conjugate, we have
\begin{equation}
   \label{eq:Eq.F7}
 (e_2^\ast (e_1+e_2 x_1+e_3 x_2+e_4 x_3+e_5 x_4+e_6 x_5+e_7 x_6))_3=0.
\end{equation}
This can be written as
\begin{equation}
\begin{split}
   \label{eq:Eq.F8}
&(e_2^\ast e_2 )_3 x_1+(e_2^\ast e_3 )_3 x_2+(e_2^\ast e_4 )_3 x_3+(e_2^\ast e_5 )_3 x_4\\
&+(e_2^\ast e_6 )_3 x_5+(e_2^\ast e_7 )_3 x_6=-(e_2^\ast e_1 )_3
\end{split}
\end{equation}
If the lowest power term in $e_3$ is 5th power, using similar method we minimize the 5th power terms in $e'_1$ by $x_2$, and find
\begin{equation}
\begin{split}
   \label{eq:Eq.F9}
&(e_3^\ast e_2 )_5 x_1+(e_3^\ast e_3 )_5 x_2+(e_3^\ast e_4 )_5 x_3+(e_3^\ast e_5 )_5 x_4\\&+(e_3^\ast e_6 )_5 x_5+(e_3^\ast e_7 )_5 x_6=-(e_3^\ast e_1 )_5
\end{split}
\end{equation}
Assuming the lowest power terms in the basis are power of 1,3,5,7,3,5,7 (Table II) respectively, we can continue the above procedure and find the following matrix equation:
\begin{equation}
   \label{eq:Eq.F10}
   \begin{split}
 &\begin{bmatrix}
(e_2^\ast e_2 )_3&(e_2^\ast e_3 )_3&(e_2^\ast e_4 )_3&...&&(e_2^\ast e_7 )_3\\(e_3^\ast e_2 )_5&(e_3^\ast e_3 )_5&(e_3^\ast e_4 )_5&...&&(e_3^\ast e_7 )_5\\(e_4^\ast e_2 )_7&(e_4^\ast e_3 )_7&(e_4^\ast e_4 )_7&...&&(e_4^\ast e_7 )_7\\...&...&...&&&...\\(e_6^\ast e_2 )_5&(e_6^\ast e_3 )_5&(e_6^\ast e_4 )_5&&&(e_6^\ast e_7 )_5\\(e_7^\ast e_2 )_7&(e_7^\ast e_3 )_7&(e_7^\ast e_4 )_7&...&&(e_7^\ast e_7 )_7
 \end{bmatrix}
  \begin{bmatrix}
x_1\\x_2\\x_3\\...\\x_5\\x_6
 \end{bmatrix}\\
 &=-
   \begin{bmatrix}
(e_2^* e_1 )_3\\(e_3^* e_1 )_5\\(e_4^* e_1 )_7\\(e_5^* e_1 )_3\\...\\(e_7^* e_1 )_7
 \end{bmatrix}
 \end{split}
\end{equation}
By Jordan decomposition the vectors $e_j$ in u (u is the column of basis as given by Eq.(F.5)) are known, the Eq.\,\eqref{eq:Eq.F10} can be solved to determine $x_j$ to establish the new basis $e'_j$. The basis of Jordan form is uniquely determined now.




%
\vspace{0cm}

\end{document}